\documentclass{aa}  

\usepackage{graphicx}
\usepackage{txfonts}
\usepackage{amsmath,amsfonts,amssymb}
\usepackage{textcomp,gensymb}
\usepackage[T1]{fontenc}
\usepackage{aecompl}
\usepackage{ulem}
\usepackage{aas_macros}
\usepackage[dvipsnames]{xcolor}
\usepackage{color}
\usepackage{times}
\usepackage{subfig}
\usepackage{stfloats}
\usepackage[utf8]{inputenc}
\usepackage{hyperref}
\usepackage{multirow}
\usepackage{natbib}
\usepackage{placeins}

\newcommand{\mdy}[1]{#1}

\newcommand{\rmvd}[1]{}

\begin{document} 

   \title{Probing dust and grain growth in the optically thick circumbinary ring of V892~Tau}
   \author{Antoine Alaguero \inst{1}
          \and
          Fran\c cois Ménard \inst{1}
          \and
          Nicol\'as Cuello \inst{1}
          \and
          \'Alvaro Ribas \inst{2}
          \and
          Elena Viscardi \inst{3}
          \and
          Enrique Mac\'ias \inst{3}
          \and\\
           Miguel Vioque \inst{3}    
          \and
          James Miley \inst{4,5,6}
          }

   \institute{Univ. Grenoble Alpes, CNRS, IPAG, 38000 Grenoble, France\\
              \email{antoine.alaguero@univ-grenoble-alpes.fr}  
         \and
             Institute of Astronomy, University of Cambridge, Madingley Road, Cambridge, CB3 0HA, UK 
         \and
             European Southern Observatory, Karl-Schwarzschild-Str. 2, 85748 Garching bei München, Germany 
         \and
            Departamento de Física, Universidad de Santiago de Chile, Av. Victor Jara 3659, Santiago, Chile  
         \and
            Millennium Nucleus on Young Exoplanets and their Moons (YEMS), Chile 
        \and
            Center for Interdisciplinary Research in Astrophysics and Space Exploration (CIRAS), Universidad de Santiago, Chile 
             }

   \date{Received xxx; Accepted xxx}

 
  \abstract
   {A considerable proportion of young stars belong to multiple star systems. Constraining the planet formation processes in multiple stellar systems is then key to understand the global exoplanet population.}
   {This study focuses on investigating the dust reservoir within the triple system V892~Tau. Our objective is to establish constraints on the properties and characteristics of the dust present in the system’s circumbinary ring.}
   {Based on archival ALMA and VLA data from $0.9$~mm to $9.8$~mm, we present a multi-wavelength analysis of the ring of V892~Tau. We first studied the spatial variation of the spectral index, before employing 3D full radiative transfer calculations to constrain the ring's geometry and the radial dependence of the dust grain properties.
   }
   {Spectral indices are consistent with non-dust emission in the vicinity of the central binary, and with dust emission in the ring likely remaining optically thick up to $3.0$~mm. 
   Our radiative transfer analysis supports these interpretations, yielding a model that reproduces the observed intensities within the $1\sigma$ uncertainties across all wavelengths. The resulting dust surface density and temperature profiles both decrease with increasing radius, and are in agreement with values reported in the literature. Maximum grain sizes are constrained to $0.2$~cm, with a size distribution power-law index $-3.5$. These results imply that the dust grain fragmentation velocity does not exceed $8$ m s$^{-1}$.}
   {Whilst our results suggest dust trapping at the cavity edge, they also suggest that tidal perturbations triggered by the central binary limit grain growth within the ring. This highlights the need to further constrain planet formation efficiency in multiple stellar systems, a goal that may be advanced by applying the methodology of this work to a wider sample of systems.
   }

   \keywords{protoplanetary discs --- 
             radiative transfer --- 
             methods: observational --- 
             stars: individual: V892~Tau ---
             binaries: general
               }

   \maketitle
%
\section{Introduction}
\label{sec:intro}

A growing number of planets are detected around binary stars thanks to recent surveys like BEBOP \citep{BEBOP}, even though it was thought that binary systems would hinder planet formation \citep{Hatzes2016}. Understanding the formation of circumbinary planets constitutes a key step toward a global understanding of the demographics of exoplanets, especially given that stars commonly form in multiple systems \citep{Offner+2022}. To this end, particular attention should be given to circumbinary discs, as they constitute the natural progenitors of circumbinary planets. Circumbinary discs are complex dynamical objects shaped by gravitational interactions with the binary components, which can further affect planet formation \citep{Cuello+2025}. For example, non-zero eccentricity developed by circumbinary discs can result in parametric instabilities \citep{Papaloizou2005a,BarkerOgilvie2014,Ragusa+2020}. This generates turbulence close to the inner binary, mitigating vertical settling and the efficiency of subsequent planet formation processes, such as streaming instability or pebble accretion \citep{Pierens+2021}. More broadly, elevated levels of gas turbulence are known to promote dust grain fragmentation \citep{StepinskiValageas1997}. Consequently, constraining dust grain sizes in circumbinary discs may offer insight into whether binary stars indeed enhance disc turbulence. Nevertheless, circumbinary discs are also known to form high dust-to-gas ratio clumps in their inner regions, particularly around unequal-mass and eccentric binaries \citep{Poblete+2019}. It suggests that these environments may be more conductive for planetesimal formation than previously thought .

Circumbinary discs often exhibit various morphological features induced by the gravitational torque of their host binary stars \citep[e.g.][]{Calcino+2023}. Numerical studies have shown that these substructures take the form of spiral arms \citep{Thun+2017,Penzlin+2024}, a central cavity with fast flows of gas \citep{MirandaLai2015}, or large-scale azimuthal asymmetries at the edge of the cavity \citep{Ragusa+2017}. These high-density regions may act as efficient traps for dust particles, potentially enhancing dust growth locally. Circumbinary discs are often found misaligned with respect to the orbital plane of the binary, which might lead to an efficient creation of dust traps \citep{Czekala+2019,Aly+2024,Smallwood+2024b}.
Establishing the properties of dust grains in the substructures of circumbinary discs and investigating their capacity at promoting dust growth are of primordial importance to understand the formation pathways of circumbinary planets.

The Atacama Large Millimeter/submillimeter Array (ALMA) has been widely used to constrain the properties of dust grains. One technique used to study grain growth probes the polarisation of the emission caused by the self-scattering of dust grains \citep[e.g.][]{YangLi2020}. Another option is to use a set of observations at several wavelengths to fit the spectral energy distribution (SED) of the dust continuum emission \citep[e.g.][]{Kim+2019}. Both techniques provide detailed insights on the dust grain population characteristics such as size \citep{Kataoka+2015,Macias+2021}, porosity \citep{Guidi+2022,Zhang+2023} and composition \citep{Tazaki+2019,Hu+2024}. 
However the impact of stellar multiplicity on the characteristics of dust grains remains unclear. The work of \cite{Sierra+2024} focused on cavity-hosting discs, some of which are part of multiple systems. It showed no particular correlation between grain size and number of stars in the system, but suggested that dust traps at the cavity edge of massive discs could be triggered by companions. Some of these companions could hide in the cavity of lopsided discs \citep{Ragusa+2025}, like AB Aur \citep{Tang+2017,Poblete+2020}, IRS 48 \citep{Calcino+2019,Yang+2023}, or ISO-Oph 2 \citep{Cieza+2021}. In these three asymmetric discs, the regions of highest density do not consistently correspond to local maxima in dust grain size, further questioning the impact of stellar multiplicity on dust grain growth \citep{Ohashi+2020,Casassus+2023,Riviere-Marichalar+2024}. 
A comprehensive understanding of dust grain growth within the substructures of discs in multiple stellar systems then relies on systems where the disc dynamics have been thoroughly characterized, and for which the substructures have been directly linked to ongoing interactions with companion stars. One such system that offers an ideal case study is V892~Tau, one of the closest Herbig stars \citep{The+1994}. The circumbinary disc of V892~Tau has been extensively targetted by recent observations, which has allowed to constrain the orbital parameters of the three stars and to highlight their interactions with the disc \citep{Long+2021,Vides+2023,Alaguero+2024}.


Here we present a multi-wavelength study of the circumbinary disc of the triple system V892~Tau. We describe the target and the observations supporting our analysis in Section \ref{sec:obs}. In that same Section, we derive preliminary results on the emission of V892~Tau through the analysis of radial profiles and spectral index considerations. We then carry out a more in-depth modelling of the system based on radiative transfer calculations in Section \ref{sec:rt_models}. Finally, we discuss our results in Section \ref{sec:discussion} and draw our conclusions in Section \ref{sec:conclusion}.

\section{Observations}
\label{sec:obs}

\subsection{Target description}
\label{subsec:target}

V892~Tau (J2000 04h18m40.62s +28d19m15.16s) is a triple stellar system in the Taurus star forming region located at a distance of $134.5\pm1.5$ pc \citep{GaiaDR3}. V892~Tau is composed by an equal mass ratio binary star of total mass $6.1\pm0.2$ M$_{\odot}$ \citep{Vides+2023}. The stars are surrounded by a circumbinary disc, beyond which a low-mass companion star is located $4\arcsec$ to the NE from the binary \citep{Skinner1993,Monnier+2008,Vides+2023}. 
V892~Tau, being heavily obscured at optical wavelengths \citep{HerczegHillenbrand2014}, has long been the focus of extensive interferometric studies \citep[on top of previously cited works, ][]{Beckwith+1990,Haas+1997,Liu+2005,Smith+2005,Panic+2009,Hamidouche+2010}. More recently \cite{Long+2021} presented a complete study of V892~Tau using data from $1.3$~mm to $9.8$~mm. Their observations revealed a dust ring with a peak intensity radius of $\sim28$ au and gas emission encircling $90\%$ of the total flux at a radius of $\sim195$ au. Interestingly, the disc is misaligned with respect to the central binary by $\sim8\degree$ and shows non-Keplerian structures. \cite{Alaguero+2024} later identified spiral arms in the gaseous disc and suggested ongoing interactions with V892~Tau~NE to explain the observed structures and misalignments. With a semi-major axis ratio between the inner binary and the outer binary of $a_{out}/a_{in}\approx 75$ \citep{Vides+2023,Alaguero+2024}, V892~Tau's architecture is characteristic of hierarchical multiple stellar systems and provide an ideal laboratory for investigating the influence of stellar multiplicity on planet formation processes.

\begin{figure*}
\centering
\begin{center}
    \includegraphics[width=\textwidth, trim={0cm 2cm 0cm 0cm},clip]{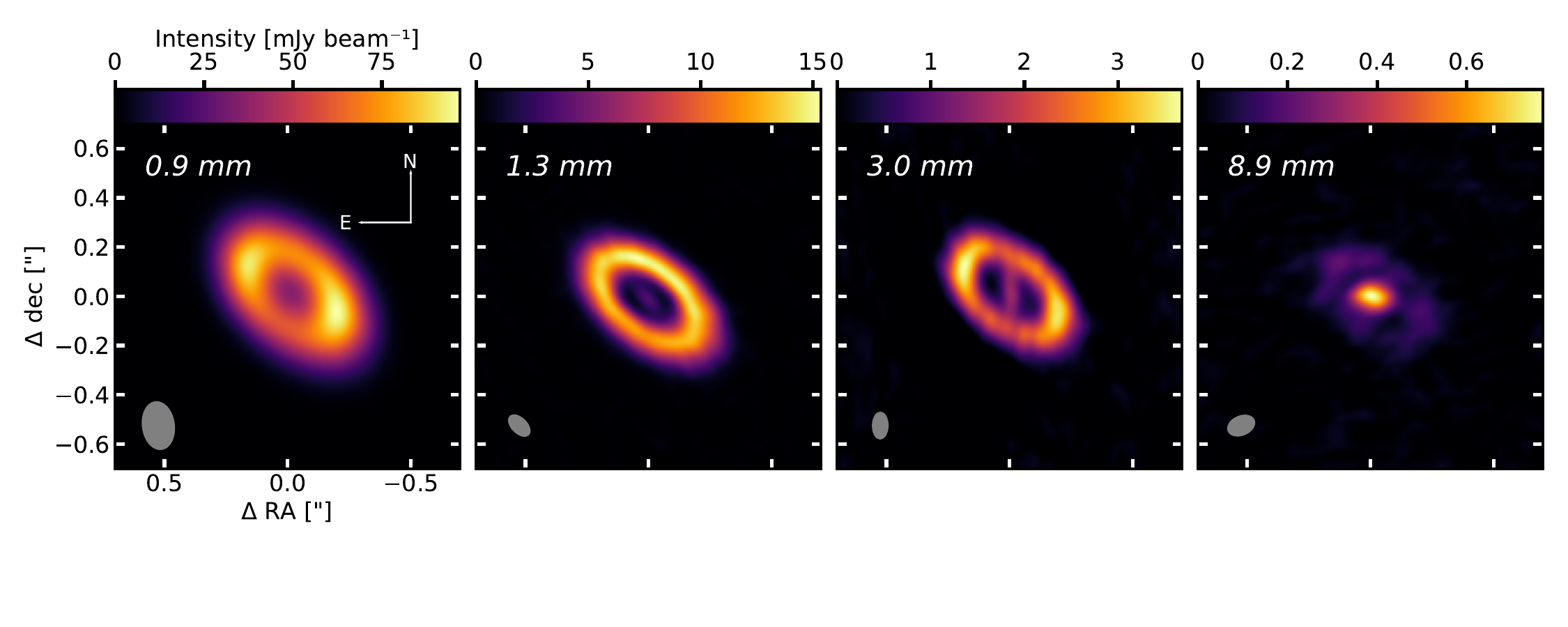}     
    \caption{From left to right : Continuum maps of V892~Tau from ALMA observations at $0.9$ mm, $1.3$ mm, and $3.1$ mm, and from combined VLA observations at $8.0$ mm and $9.8$ mm. The size of the synthesized beam is represented by the grey ellipse at the bottom left of each image. Orientation of the sky plane is indicated in the leftmost panel.}
    \label{fig:obs_gallery}
\end{center}
\end{figure*}

We use in this work ALMA data of V892~Tau at three different wavelengths, and Karl G. Jansky Very Large Array (VLA) observations at two different wavelengths. We describe the ALMA and VLA observations in the following Section \ref{subsec:ALMA_obs} and \ref{subsec:VLA_obs}, respectively.
Figure \ref{fig:obs_gallery} shows the images obtained after the reduction process while Table \ref{table:comb_obs} summarizes their properties.

\begin{table*}
    \centering
    \caption{Properties of the images obtained from the reduced datasets.}
 \begin{tabular}{c c c c c c c c c} 
Wavelength & Beamsize (beam PA) & Flux \tablefootmark{1} & RMS & $R_{\text{cav}}$ & $R_{\text{peak}}$ & $R_{90\%}$ & i & PA\\ [0.5ex] 
(mm) & $(\arcsec)$ & (mJy) & (mJy\,beam$^{-1}$) & $(\arcsec)$ \tablefootmark{2} & $(\arcsec)$ & $(\arcsec)$ & ($\degree$) \tablefootmark{3} & ($\degree$) \tablefootmark{3} \\
 \hline \hline
$9.8$\tablefootmark{4} & $0.11 \times 0.07$ ($-6.3\degree$) & $1.8\pm0.3$ & $0.025$ & / & / & / & / & / \\
$8.0$\tablefootmark{4} & $0.12 \times 0.08$ ($-7.0\degree$) & $3.1\pm0.4$ & $0.035$ & / & / & / & / & / \\
$8.0+9.8$\tablefootmark{5} & $0.11 \times 0.08$ ($-70.0\degree$)  & $2.0\pm0.1$ & $0.015$ & $0.15\pm0.02$ & $0.21\pm0.01$ & $0.41\pm0.01$ & / & / \\
$3.0$ & $0.11 \times 0.06$ ($-0.7\degree$) & $51.6\pm0.1$ & $0.095$ & $0.15\pm0.10$ & $0.23\pm0.01$ & $0.37\pm0.01$ & $55.2$ & $53.3$ \\
$1.3$ & $0.11 \times 0.06$ ($45.2\degree$) & $297.1\pm0.1$ & $0.089$ & $0.14\pm0.03$ & $0.22\pm0.01$ & $0.38\pm0.01$ & $55.3$ & $52.3$ \\
$0.9$ & $0.19 \times 0.13$ ($7.8\degree$) & $677.8\pm0.1$ & $0.120$ & $0.05\pm0.01$ & $0.21\pm0.01$ & $0.43\pm0.01$ & $56.0$ & $52.6$ \\
\end{tabular}
\tablefoot{
\tablefootmark{1}{The flux uncertainty for the ALMA images was measured by multiplying the noise level by the square root of the number of pixels in the integrated area.} \\
\tablefootmark{2}{$R_{\text{cav}}$ is measured as the radius at which the emission reaches half of its peak value. The error is determined by the difference between this half-peak radius and the radius at which the emission reaches a quarter of its peak value. } \\
\tablefootmark{3}{The inclination and Position Angle (PA) values are derived from a geometric fit to the visibility data done with \textit{frank} \citep{Jennings+2020}.} \\
\tablefootmark{4}{Data presented in \cite{Long+2021}, with the corresponding fluxes measured after subtraction of a binary star model. 
}\\
\tablefootmark{5}{Dataset created by combining the datasets at $8.0$ mm and $9.8$ mm after the subtraction of a binary star model.}
}
\label{table:comb_obs}
\end{table*}

\subsection{ALMA Observations}
\label{subsec:ALMA_obs}

The Band $3$ data (of central wavelength $\lambda_0=3.0$ mm) were acquired in the context of the ALMA programme 2016.1.01042.S (PI: C. Chandler). The Band $6$ data ($\lambda_0=1.3$ mm) were acquired in the context of the ALMA programmes 2013.1.00498.S (PI: L. Pérez) and 2021.1.01137.S (PI: J. Miley). The Band $7$ data ($\lambda_0=0.9$ mm) were acquired in the context of the ALMA programme 2017.1.00470.S (PI: L. Looney). All datasets are publicly available on the ALMA Science Archive\footnote{\url{https://almascience.eso.org/aq/}}. The standard ALMA calibration was applied and the calibrated data retrieved using the \textit{CalMS} service from the European ALMA Regional Center.

The Band $3, 6$ and $7$ data are each composed of $3$, $3$, and $4$ execution blocks (EB hereafter) respectively. For each EB, the channels including spectral lines were flagged before all the channels were combined together to create continuum datasets.
We used the method described in \cite{Alaguero+2024} (their Appendix A) to self-calibrate and combine the EBs at each wavelength. 
At a given wavelength, the individual EBs are first iteratively self-calibrated in phase. Calibration tables were produced through the \textit{gaincal} task, with decreasing solution time interval through the iterations down to the integration time.
If the Signal to Noise Ratio (S/N) improved compared to the previous iteration, the calibration tables found were applied to the data with \textit{applycal} in \textit{calonly} mode to ensure no data was flagged. 
Disc flux, peak value, and noise level of the background were measured at each step to assess the S/N. These first rounds of self-calibration resulted in the peak S/N increasing by $43 \%$, $21 \%$, and $514 \%$ for EBs of Band $3$, $6$, and $7$, respectively. Then, for each Band, the self-calibrated EBs were centred at the centre of the cavity. Except in the case of the Band $6$ data where EBs were rescaled \citep{Alaguero+2024}, the total fluxes of the datasets were consistent within $5\%$ and no flux rescaling were performed. The EBs at each wavelength were then imaged together using \textit{tclean} to create a common model. From this model, the EBs were jointly self-calibrated in phase a second time. Calibration from a common model increased the S/N by $5 \%$, $9 \%$, and $30 \%$ for EBs in Band $3$, $6$, and $7$, respectively. At the end, the EBs at each wavelength were imaged together with \textit{tclean} using a Briggs weighting with a robust parameter of $-0.5$ to enhance angular resolution. The final images are shown in Figure \ref{fig:obs_gallery} while the corresponding fluxes, RMS, and synthesized beams at each wavelength can be found in Table \ref{table:comb_obs}.

\subsection{VLA Observations}
\label{subsec:VLA_obs}

We used the VLA data published in \cite{Long+2021} and we refer the reader to their work for an extensive description of these observations. 
The observations were performed in the K$_{a}$ band, at $30.5$ GHz ($\lambda_0=9.8$ mm) and $37.5$ GHz ($\lambda_0=8.0$ mm), in the A, B and C configurations of the VLA. In an attempt to enhance the S/N, the datasets at $30.5$ GHz and $37.5$ GHz were combined and imaged together using \textit{tclean} with a Briggs weighting parametrized by a robust parameter of $0.5$ to further enhance sensitivity. As shown by Table \ref{table:comb_obs}, it resulted in similar synthesized beam sizes than for the ALMA Band $3$ and Band $6$ images. 
To mitigate contamination from non-dust emission, we removed the central binary emission using methods described in Appendix \ref{app : free-free}. This procedure was applied to the datasets at $30.5$~GHz and $37.5$~GHz, prior to their combination and subsequent imaging as described earlier in this Section.

\subsection{Disc morphology and radial profiles}
\label{subsec:oui}

The ALMA images display a bright ring encircling fainter inner regions. Central emission at $1.3$~mm and $3.0$~mm appears to be compatible with a point source, but is more diffuse at $0.9$~mm because of a lower angular resolution.
At $0.9$~mm and $3.0$~mm, the convolution with a N-S aligned beam makes the ring brighter at its NE and SW edges. On the $3.0$ mm emission map, bridges connect the inner disc to the ring. Because the bridges has the same orientation as the synthesizing beam, these features likely are observational artifacts resulting from beam convolution. As noted by \cite{Long+2021} at $1.3$ mm, the NE side of the disc is brighter than the SE. Peak pixel values along the disc minor axis show are brighter in the NW compared to the SE by approximately $22\%$, $22\%$ and $13\%$ at $0.9$~mm, $1.3$~mm, and $3.0$~mm, respectively. \cite{Ribas+2024} attribute this to optically thick emission from the inner rim, visible only on the far side.

The central emission dominates the VLA images, where sparse emission is observed at the location of the ALMA ring. The emission associated with the ring is detected at $3-5$ S/N levels with a clumpy structure attributed to noise fluctuations \citep{Macias+2017,Long+2021}. On Figure \ref{fig:obs_gallery}, the two central stars are marginally resolved but full resolution can be achieved when considering the VLA longest baselines solely \citep{Long+2021}.

We estimated the extent and position of the ring at each wavelength based on radial profiles. Assuming the centre of the cavity to be the centre of the disc, an inclination of $i=55.5\degree$, and a position angle (PA, defined East of North) PA $=52.7\degree$, the deprojected radial profiles were computed from the images by azimuthally averaging the intensity in concentric ellipses (see Section \ref{subsec:geo_fit} for the determination of the disc geometry). The average intensity $I_{\nu,j}$ was computed in each radial bin $j$ alongside the associated uncertainty $\sigma_{I_{\nu}, i}$:
\begin{equation}
\label{eq:rad_error}
\sigma_{I_{\nu}, j} = \frac{\sigma_{j}}{\sqrt{A_j / A_{beam}} } \,,
\end{equation}
with $\sigma_{j}$ the standard deviation in the radial bin $j$, A$_j$ the area of the radial bin and A$_{beam}$ the beam area.

\begin{figure*}
\centering
\begin{center}
    \includegraphics[width=\textwidth, trim={0cm 0cm 0cm 0cm},clip]{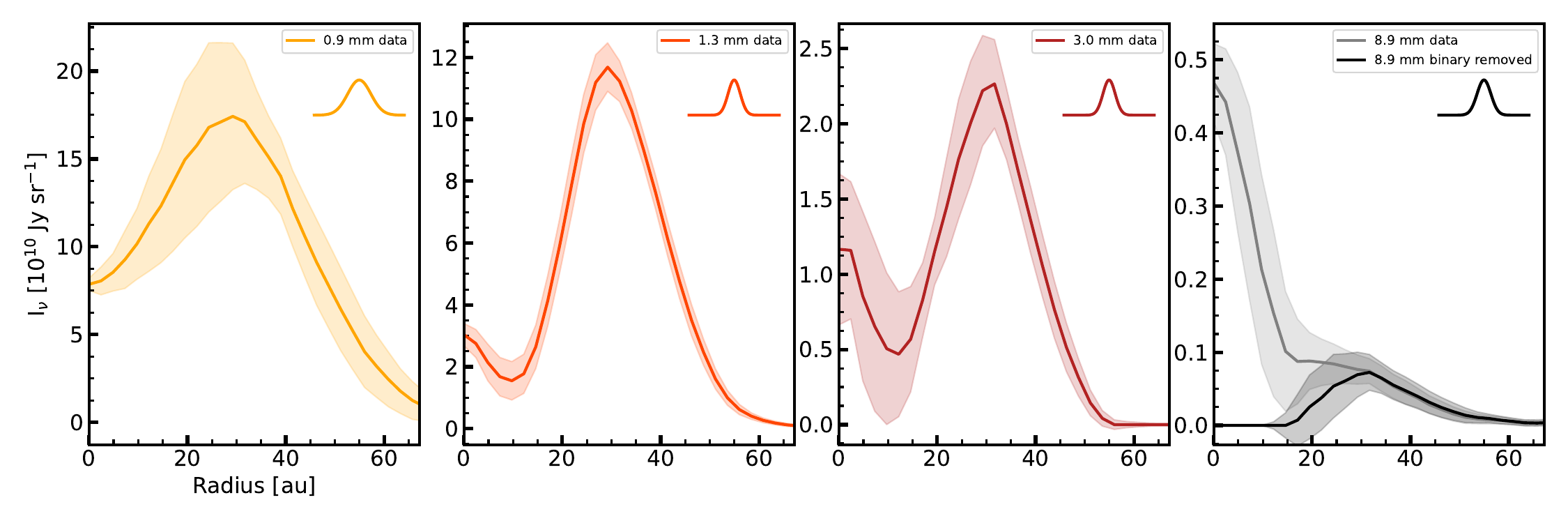}     
    \caption{
    Deprojected radial profiles computed from the images shown in Figure \ref{fig:obs_gallery}, respectively at $0.9$~mm, $1.3$~mm, $3.0$~mm, and combined observations of effective wavelength $8.9$~mm from left to right. The shaded area in each panel indicates the uncertainty computed following Equation \ref{eq:rad_error}. At $8.9$~mm the radial profile of the binary-subtracted image is also shown. The average geometrical size of the beam is plotted in the top right of each panel.}
    \label{fig:rad_prof}
\end{center}
\end{figure*}

Figure \ref{fig:rad_prof} shows the radial profiles obtained from the data images and their associated errors. 
The radial profiles of the ALMA observations peak close to $0.209\pm0.001\arcsec$, which is the position of the ring measured by \cite{Long+2021}. Signal is observed in the innermost regions, which highlights the presence of emission at the centre of the system. The uncertainties are larger at the peak radius, reflecting significant intensity variations along the azimuth. 
At $3.0$~mm, beam convolution connects the central emission to the disc, resulting in larger intensity scatter in the cavity and thus in greater uncertainties.
The radial profile of the VLA observations peaks at the centre of the system. A local plateau is seen at approximately $31$ au ($0.23\arcsec$), which is consistent with the peak position $R_{\text{peak}}$ of the dust ring seen with ALMA. Once the central emission is removed, $R_{\text{peak}}$ becomes consistent between the ALMA and VLA observations. Beyond the emission peak, the binary-subtracted radial profile matches the original VLA data. 

From the ALMA intensity radial profiles, we measured the cavity radius $R_{\text{cav}}$, the peak intensity radius $R_{\text{peak}}$, and the disc radius $R_{90\%}$. Their values are reported in Table~\ref{table:comb_obs}. $R_{\text{peak}}$ traces the position of the ring, which is comparable between all the observations. $R_{\text{cav}}$ is defined as the radius at which the intensity reaches half of its peak value, excluding central emission from the calculation if present. The uncertainty on $R_{\text{cav}}$ is taken as the difference with the radius at which the emission reaches a quarter of its peak value. $R_{\text{cav}}$ is expected to remain almost constant across mm wavelengths, which trace the emission of mm-sized grains trapped at the cavity edge \citep{Pinilla+2012}. At $1.3$~mm and $3.0$~mm, the measured values of $R_{\text{cav}}$ are consistent with one another and support this scenario. At $0.9$~mm, however, beam dilution leads to an underestimation of $R_{\text{cav}}$. Finally, $R_{90\%}$ is defined as the radius encircling $90\%$ of the total flux. Radial distances within $0.08\arcsec$ are excluded from the calculation of $R_{90\%}$ to ensure a reliable estimate of the ring’s characteristic radius. Decreasing values of $R_{90\%}$ are measured with increasing wavelength. However this decrease is insufficient to be attributed to radial drift \citep{Rosotti+2019b}, and is likely due to a larger beam size at $0.9$ mm. 
$R_{\text{cav}}$, $R_{\text{peak}}$, and $R_{90\%}$ cannot be robustly estimated for the VLA datasets individually due to the low signal levels. However, the S/N is higher in the image combining the VLA datasets at $8.0$~mm and $9.8$~mm. We obtained for this combined dataset a cavity radius $R_{\text{cav}}=0.15\pm0.02\arcsec$, a peak intensity radius $R_{\text{peak}}=0.21\pm0.01\arcsec$, and disc radius $R_{90\%}=0.41\pm0.01\arcsec$. These values are in good agreement with the ALMA observations. 

\subsection{Spectral index}
\label{subsec:sp_ind}

The spectral index $\alpha$ is defined as:
\begin{equation}
    \label{eq:flux_spectral_index}
    \alpha = \frac{{\rm d}\log F_{\nu}}{{\rm d}\log\nu} \, .
\end{equation}
In the millimetre regime, a good approximation for the absorption opacities of dust grains is $\kappa_{\text{abs}} \propto \nu\,^{\beta}$ with $\beta$ the spectral index of the dust opacity \citep{Draine2006}. Neglecting scattering and assuming optically thin emission in the Rayleigh-Jeans regime, the flux verifies $F_{\nu}\propto\kappa_{\text{abs}}\nu^2$. Under these assumptions, $\alpha$ is linked to the opacity by the relation $\alpha = \beta+2$. Since $\beta$ depends on the dust composition and size \citep[e.g.][]{Pollack+1994,Draine2006}, measuring $\beta$ allows to probe the dust properties in the disc \citep{Testi+2014}. However, optically thick regions of the disc may contribute to the measured $\alpha$, marking significant deviations from the previous relation \citep{Beckwith+1990}. In case of optically thick emission in the Rayleigh-Jeans regime, we have indeed $F_{\nu}\propto\nu^2$, implying  $\alpha=2$ regardless of the dust composition, size, and shape. Non-dust emission can be produced through processes related to ionized plasma like jets \citep[e.g.][]{Reynolds1986}, disc winds \citep[e.g.][]{Pascucci+2012}, or stellar magnetic activity \citep[e.g.][]{Andre1996}. These processes typically result in values of  spectral index in the range $-0.1<\alpha<1.5$ \citep[e.g.][]{Rota+2024}.
Measuring $\alpha$ then allows to determine the nature of the emission between non-dust processes and dust thermal emission. For dust thermal emission, further distinction can be made between optically thin and optically thick emission. In the case of optically thin emission, $\alpha$ is a direct probe of the dust opacities.
Deviations from the Rayleigh-Jeans regime or additional physics like dust self-scattering may alter the analysis, which then requires a cautious interpretation \citep[e.g.][]{Wilner+2005,Zhu+2019}.


\subsubsection{Integrated spectral index}
\label{subsubsec:sp_ind_int}

We computed the flux spectral index $\alpha$ of V892~Tau using the ALMA self-calibrated observations described in Section \ref{subsec:ALMA_obs} and the VLA observations presented in Section \ref{subsec:VLA_obs} and \cite{Long+2021}.
To do this, we separated the flux of the central regions from the one of the outer disc. To measure the flux at $1.3$~mm and $3.0$~mm, a 2D Gaussian was fitted to the central emission using the \textit{imfit} task. At $0.9$~mm, the central emission is diluted which makes the measurement less straightforward. We instead considered the inner flux as the flux inside $R_{\text{cav, 0.9mm}} = 0.05\arcsec$ in the \textit{tclean} model. The flux uncertainty for the ALMA images was measured by multiplying the noise level by the square root of the number of pixels in the integrated area. The flux uncertainty also included calibration errors, taken as $5\%$ of the flux at $1.3$~mm and $3.0$~mm and as $10\%$ of the flux at $0.9$~mm \citep{ALMAusermanual} and VLA wavelengths\footnote{\url{https://science.nrao.edu/facilities/vla/docs/manuals/oss/performance/fdscale}}.
On the VLA observations, we assumed the inner flux to be the total flux of the two point source model fitted to the data by \cite{Long+2021} (see also Appendix \ref{app : free-free}).
This allowed to derive the flux value of the outer dusty disc following the formula:
\begin{equation}
F_{out} = F_{tot} - F_{in}
\end{equation}
with $F_{tot}$ being the total flux, $F_{out}$ the flux of the outer disc, and $F_{in}$ the flux of the inner disc. Table \ref{table:fluxes} lists the fluxes of the inner and outer discs at each wavelength.

\begin{table}
    \centering
    \caption{Fluxes of the inner disc and the outer disc at each wavelength.}
 \begin{tabular}{c c c c} 
Wavelength & Frequency & $F_{in}$ & $F_{out}$\\ [0.5ex] 
(mm) & (GHz) & (mJy) & (mJy)\\
 \hline \hline
$9.8$\tablefootmark{1} & $30.5$ & $1.2\pm0.1$ & $1.8\pm0.3$ \\
$8.0$\tablefootmark{1} & $37.5$ & $1.2\pm0.1$  & $3.1\pm0.5$ \\
$3.0$ & $100$ & $0.8\pm0.9$ & $50.8\pm2.5$  \\
$1.3$ & $224$ & $3.1\pm0.8$  & $294.3\pm14.7$ \\
$0.9$ & $315$ & $5.3\pm1.1$\tablefootmark{2} & $672.5\pm67.2$ \\
\end{tabular}
\tablefoot{$F_{in}$ and $F_{out}$ correspond to the flux of the inner regions and of the outer ring, respectively. The uncertainties take into account the calibration errors of $5\%$ at $1.3$~mm and $3.0$~mm, and of $10\%$ at $0.9$~mm, $8.0$~mm, and $9.8$~mm. \\
\tablefootmark{1}{Data presented in \cite{Long+2021}, with the corresponding flux of the outer disc measured after the subtraction of binary model and the flux of the inner disc taken as the total flux of that model.} \\
\tablefootmark{2}{This value was measured by integrating the intensity within $R_{\text{cav, 0.9mm}} = 0.05\arcsec$ in the \textit{tclean} model.}}
\label{table:fluxes}
\end{table}

From the flux values, we measured the spectral indices of the inner disc $\alpha_{\text{in}}$ and of the outer disc $\alpha_{\text{out}}$ using Equation \ref{eq:flux_spectral_index}.  
For a better match to the data, $\alpha$ was calculated separately between $0.9$~mm and $3.0$~mm, and between $3.0$~mm and $9.8$~mm. 

\begin{figure}
\centering
\begin{center}
    \includegraphics[width=\columnwidth, trim={0cm 0cm 0cm 0cm},clip]{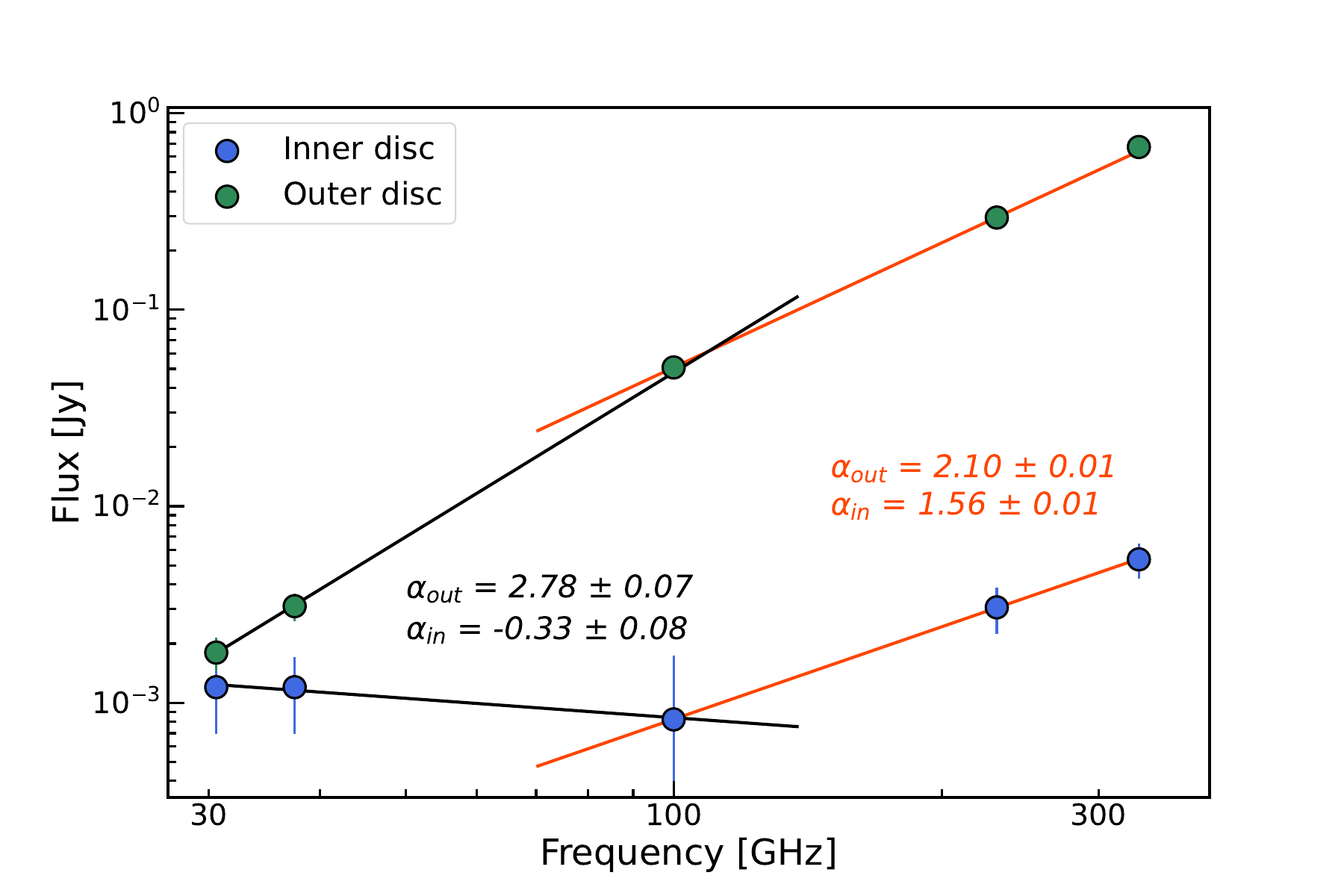}     
    \caption{SED of V892~Tau measured from ALMA and VLA observations. The fluxes of the inner and outer discs are represented by the blue and green points, respectively. Our best-fit power-laws and their corresponding spectral index are represented in plain lines in orange between $0.9$~mm and $3.0$~mm (\textit{i.e.} $100-333$~GHz), and in black between $3.0$~mm and $9.8$~mm (\textit{i.e.} $30.5-100$~GHz). $\alpha_{\text{in}}$ and $\alpha_{\text{out}}$ are the spectral indices of the inner and outer discs respectively. The error-bars for the outer disc are smaller than the markers at wavelengths shorter than $3.0$~mm.}
    \label{fig:sp_ind_integrated}
\end{center}
\end{figure}

Figure \ref{fig:sp_ind_integrated} shows the SED of the inner and outer discs and their respective spectral index.
In the inner disc, we find $\alpha_{\text{in}} = 1.55 \pm 0.01$ between $0.9$~mm and $3.0$~mm, and $\alpha_{\text{in}} = -0.33 \pm 0.08$ between $3.0$~mm and $9.8$~mm. 
We measure $\alpha_{\text{out}} = 2.10 \pm 0.01$ between $0.9$~mm and $3.0$~mm, and $\alpha_{\text{out}} = 2.78 \pm 0.07$ between $3.0$~mm and $9.8$~mm. These values suggest a difference in the nature of the emission between the regions in the disc as a function of wavelength. Our spectral index values are in agreement with \cite{Rota+2024}, where the authors measured $\alpha_{\text{in}}$ and $\alpha_{\text{out}}$ for a dozen of discs harbouring cavities. Between $0.9$~mm and $3.0$~mm the value of $\alpha_{\text{in}}$ is compatible with a non-dust nature of the emission, which needs to be put in the context of the X-ray activity of V892~Tau \citep{Giardino+2004}. X-ray flares in the system may be caused by stellar winds or strong stellar magnetic activity, both of which are expected to have a spectral index of $\alpha<1$ \citep{ZinneckerPreibisch1994}. In the outer disc, having $2<\alpha_{\text{out}}<3$ is consistent with dust thermal emission and in line with large surveys including V892~Tau \citep{Andrews+2013,Harrison+2024,Chung+2024,Chung+2025}. Between $0.9$~mm and $3.0$~mm, having $\alpha_{\text{out}}\approx2$ suggests optically thick emission. At longer wavelength, the increase of $\alpha_{\text{out}}$ likely indicate a decrease in optical depth, as observed in many discs already \citep[e.g.][]{Garufi+2025}. 
Between $3.0$~mm and $9.8$~mm, $\alpha_{\text{in}}$ cannot be explained by solely non-dust emission or dust thermal emission, but rather by a combination of these processes. Additionally, the change in the slope of the emission of the inner disc occurring  at $3.0$~mm likely indicate a transition in the underlying dominant emission mechanism. As a consequence, the presence of dust in the vicinity of the binary star cannot be ruled out.

\subsubsection{Spectral index maps}
\label{subsubsec:sp_ind_maps}

Since the disc is resolved in our observations, we computed spectral index maps of V892~Tau in the image plane. We computed the spectral index between two wavelengths applying Equation \ref{eq:flux_spectral_index} to each pixel of the images. Before each computation, the two corresponding images were convolved to the a common resolution using the \textit{imsmooth} CASA task. This common resolution was chosen as the largest beam size between the two images (for the individual beam sizes see Table \ref{table:comb_obs}). The images were also centred together prior to the calculations. It was done manually based on the cavity centre at $0.9$~mm, $1.3$~mm, and $3.0$~mm, and by taking the centre of the binary star model for the VLA datasets.

\begin{figure*}
\centering
\begin{center}
    \includegraphics[width=\textwidth, trim={0cm 0cm 0cm 0cm},clip]{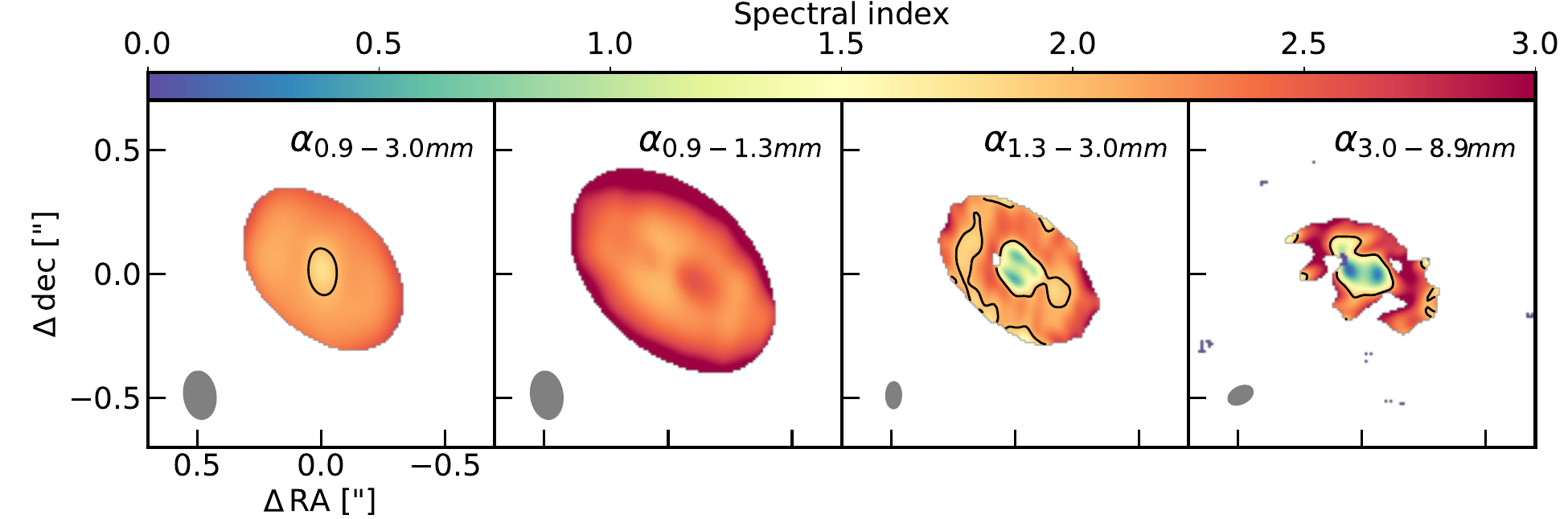}     
    \caption{
    Spectral index maps of V892~Tau. Before the calculation of the maps, a $3\sigma$ clipping has been applied to each image, which were then convolved to a common resolution. For a spectral index map between two wavelengths, the resolution was chosen accordingly to the largest beam of the two images, which is depicted at the bottom left of each panel. The black contour indicates the $\alpha=2$ threshold. 
    }
    \label{fig:gallery_alpha}
\end{center}
\end{figure*}

Figure \ref{fig:gallery_alpha} shows the resulting spectral index maps clipped at the $3\sigma$ emission level of both observations included in the calculation of $\alpha$. 
Between $0.9$~mm and $3.0$~mm, the spectral index is around $2$ in the bulk of the disc and takes lower values in the central regions. As seen in Section \ref{subsubsec:sp_ind_int}, this pattern is consistent with optically thick thermal emission of dust grains in the outer disc and with emission of non-dust processes in the environment of the stars. When the $0.9$~mm data are included, the spectral index in the inner regions is in the range $1.6-2.1$, while it is $\lesssim1$ when not. Both a difference in emission mechanisms and beam dilution of the disc signal—due to convolution with the beam at $0.9$~mm—are plausible explanations for this observation.
Between $1.3$~mm and $3.0$~mm, the spectral index is locally below $2$. This could be explained by dust self-scattering in a very optically thick region of the disc \citep{Liu2019,SierraLizano2020}.

We computed the spectral index between $3.0$~mm and the combined VLA observations, \textit{i.e.} a dataset with a representative wavelength of $8.9$~mm. The spectral index between $3.0$~mm and $8.9$~mm is found higher than in the ALMA regime, but remains lower than $3$ in the outer disc. This aligns with the optical depth decreasing in the outer disc. This decrease can be attributed to the opacity of the grains decreasing with increasing wavelength \citep[e.g.][]{WeingartnerDraine2001}. 
Dust growth and evolution models predict a rising spectral index with radius due to the drift of large grains to the inner disc, creating a decreasing gradient of maximum dust grain size with radius and thus changing the apparent disc size \citep{Rosotti+2019b,Tazzari+2021}. For further information on the radial spectral index in V892~Tau, we refer the reader to Appendix \ref{app : sp_ind_rad}.
Over the $3.0-8.9$ mm wavelength range, the spectral index is around $0.3$ in the inner regions. Optically thin free-free emission or gyro-synchrotron emission in the vicinity of the stars dominating any potential dust emission at that location could explain this observation \citep{DiFrancesco+1997,FleishmanMelnikov2003,Scaife2013}.


\section{Radiative transfer modelling}
\label{sec:rt_models}

We present in the following Section our modelling of the circumbinary ring of V892~Tau. We based our analysis on radiative transfer simulations done with the {\sc MCFOST} code \citep{Pinte+2006,Pinte+2009}. The analysis was performed in two steps. We first constrained the geometry of the system and then, with a fixed geometry, explored the dust properties in the system as a function of radius.

\subsection{Geometrical fitting}
\label{subsec:geo_fit}

The initial step in the modelling process involved determining the system's geometric configuration. We estimated the extent of the ring, defined by an inner radius $R_{in}$ and an outer radius $R_{out}$, and the ring orientation defined by an inclination $i$ and a mean PA of the disc. For this geometrical modelling, we used the ALMA data solely. Owing to the low S/N, we did not fit the disc orientation in the VLA observation. Instead, we assumed it to be identical to the value derived from the ALMA observations. 
To get the inclination and PA of the ALMA observations, we used the \textit{FitGeometryGaussian} function of \textit{frank} \citep{Jennings+2020}, which determines the disc geometry by fitting a 2D Gaussian to the visibility data. Table~\ref{table:comb_obs} gives the inclination and PA at each wavelength. Averaging over the wavelengths, the ALMA ring has an inclination of $i=55.5\degree^{+0.5}_{-0.3}$ and a position angle of PA$=52.7\degree^{+0.6}_{-0.4}$, with the uncertainty taken as the difference to the extremal values. 

Then, a first grid of {\sc MCFOST} simulations was built varying $R_{in}$ and $R_{out}$ while fixing $i$ and PA to their average values. The disc model was defined by a surface density power-law in radius $\Sigma\propto r^{p}$ between $R_{in}$ and $R_{out}$ with an exponent $p=-1$, as commonly assumed for disc models \citep{Bae+2023,Alaguero+2024}. {\sc MCFOST} requires an input dust mass to normalize the surface density profile, which was set to $M_{\text{dust}}=6\times10^{-4}$ M$_{\odot}$, in agreement with \cite{Long+2021}. The gas and dust spatial distributions were assumed to be identical. {\sc MCFOST} then relies on gas-related parameters to further define the disc. To reproduce the gas temperature profile measured from the $^{12}$CO emission by \cite{Long+2021}, we chose the disc scale height to be defined by a power-law in radius with a normalisation $H_0 = 5.5$~au at a reference radius $R_0 = 100$~au and a flaring exponent of $1.4$. Each of the stars were assumed to follow the models of \cite{Siess+2000} for a $3$ M$_{\odot}$ star at $3$ Myr, which is in agreement with the estimated masses and age of V892~Tau \citep{KucukAkkaya2010,Long+2021}. The adopted stellar parameters correspond to a surface temperature of $10 745$~K and a luminosity of 72~L${\odot}$ per star. The temperature structure were calculated from a Monte-Carlo approach using a total of $1.28\times10^7$ photon packets. Another $1.28\times10^6$ photon packets were used to produce images at the same wavelengths of the ALMA data, namely $0.9$~mm, $1.3$~mm, and $3.0$~mm. 
The dust grains were taken as compact astrosilicates \citep{WeingartnerDraine2001} with scattering properties computed according to the Mie theory. We considered $100$ logarithmically spaced bins in size, ranging from $a_{min}=0.05$~$\mu$m to $a_{max}=1$~mm. The overall distribution was normalized by integrating over all grain sizes, assuming a standard power-law exponent of $-3.5$ \citep{Mathis1977}, and across the entire grid to maintain a dust-to-gas mass ratio of $0.01$.
To reproduce $R_{\text{cav}}$ and $R_{90\%}$ (see Table \ref{table:comb_obs}), $R_{in}$ and $R_{out}$ were respectively varied between $15$~au and $29$~au, and between $40$~au and $54$~au in steps of $1$~au. Before going further, the models were scaled to the observed $F_{out}$ at each wavelength to erase eventual flux discrepancies. Following Table \ref{table:fluxes}, $F_{in}$ was added by hand on the central pixel of each image. Potential flux discrepancies will be addressed in the next Section \ref{subsec:dust_modelling}. From the computed images, we built synthetic visibility data at the same uv points than the ALMA data using the python package {\sc Galario} \citep{Tazzari+2018}. The agreement between the data and the models was then calculated using the likelihood $\mathcal{L}\propto\exp(-\chi^2_{\text{geo}}/2)$ with 
\begin{equation}
\label{eq:chi2_geo}
\chi^2_{\text{geo}} = \sum_{i} \Re (V_i - V_{i,\text{mod}})^2 * w_i \,,
\end{equation}
where $V_i$, $V_{i,\text{mod}}$, and $w_i$ are respectively the data visibility, the model visibility, and weight at the uv point $i$. $\chi^2_{\text{geo}}$ was calculated using all the visibility points of the ALMA datasets. Since $R_{in}$ and $R_{out}$ can be determined without considering the disc brightness asymmetries, the fit was performed on the real part of the visibility points.

\begin{figure}
\centering
\begin{center}
    \includegraphics[width=\columnwidth, trim={0cm 0cm 0cm 0cm},clip]{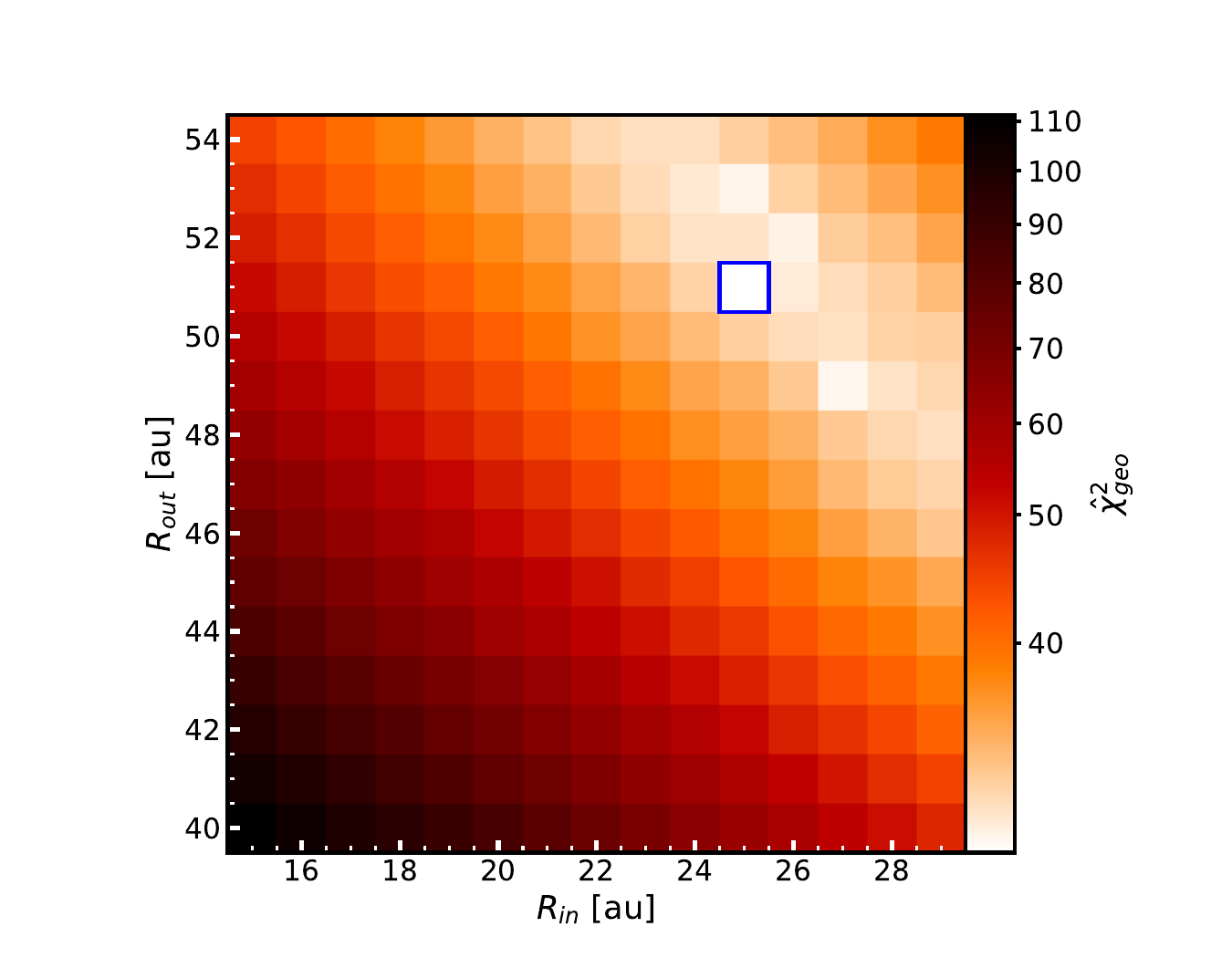}     
    \caption{Reduced $\chi^2_{\text{geo}}$ map of the geometric models tested for V892~Tau. The best-fit model is highlighted by the blue square and consists of an inner radius of $R_{in}=25$ au and an outer radius $R_{out}=51$ au. It is found with $\Hat{\chi}^2_{\text{geo}}=33$.}
    \label{fig:chi2_geo}
\end{center}
\end{figure}

The reduced $\chi^2_{\text{geo}}$ values, noted $\Hat{\chi}^2_{\text{geo}}$, are shown as a function of $R_{in}$ and $R_{out}$ on Figure \ref{fig:chi2_geo}. The best-fit model for the dust ring structure is found for $R_{in}=25^{+4}_{-2}$~au and $R_{out}=51\pm3$~au. The uncertainties were taken as the limits of the $1\sigma$ confidence interval around the best-fit values \citep{Andrae2010}. The best-fit model has a large value of $\Hat{\chi}^2_{\text{geo}}=33$, which could be improved by exploring the parameter space with a finer grid. However, as this geometric fitting step primarily served to inform subsequent, more detailed modelling—and given the strong agreement between the data and our comprehensive models—we consider the current sampling to be adequate. 

\subsection{Modelling of the dust properties}
\label{subsec:dust_modelling}

In order to model the dust properties in V892~Tau, we built a second grid of radiative transfer simulations using {\sc MCFOST}. According to the previous Section, the system's geometry was fixed to an inner radius of $R_{in}=25$ au, an outer radius of $R_{out}=51$ au, an inclination of $i=55.5\degree$ and a position angle PA$=52.7\degree$. The disc surfaces, the modelled stars and packets of photons were the same as described in the previous Section \ref{subsec:geo_fit}. \mdy{Each of the models had a surface density profile defined by a power-law of exponent $p=-1$. The surface density profile was normalised to the total dust mass, that we varied across the models.} From the flux at $1.3$~mm, the dust disc has a mass of approximately $6\times10^{-4}$ M$_{\odot}$ \citep{Long+2021}. Since this measurement is likely a lower limit due to optical depth effects \citep[e.g.][]{Liu2019}, the total dust mass was sampled between $M_{\text{dust}}=5\times10^{-5}$ M$_{\odot}$ and $M_{\text{dust}}=5\times10^{-2}$ M$_{\odot}$ with $10$ values per dex. In addition, dust grains were represented as compact spheres with the DSHARP composition \citep{Birnstiel+2018}. Using the python package {\sc dsharp$\_$opac}\footnote{\url{https://github.com/birnstiel/dsharp\_opac}}, we computed the dielectric constants of the DSHARP mixture composed by $20\%$ water ice \citep{WarrenBrandt2008}, $33\%$ silicates \citep{Draine2003}, $7\%$ troilite, and $40\%$ of organics \citep{HenningStognienko1996}. {\sc MCFOST} then computed the opacities of the input grain population using the dielectric constants. The dust grain size distribution is sampled by $100$ logarithmically spaced bins from $a_{min}=0.05$~$\mu$m to a variable $a_{max}$. Simulations were realized for $a_{max}$ located between $0.01$~cm and $100$~cm with $5$ values per dex, but $a_{max}$ values lower than $0.03$ cm were ignored to avoid degeneracies in the opacities (see opacity curves in Appendix \ref{app : opacities}). The power-law exponent $q$ of the size distribution was set equal to either $-2.5$, $-3.0$, or $-3.5$. The final parameter space was then sampled by a total of $1800$ simulations. 

From the simulations, images were computed at $0.9$~mm, $1.3$~mm, $3.0$~mm, and $8.9$~mm before being convolved to a common resolution. The limiting resolution is imposed by the $0.9$ mm observations, whilst the other observations have a better resolution (see Table \ref{table:comb_obs}). To enhance the angular resolution of the analysis, we excluded the $0.9$~mm observations from the fitting procedure.
Both data and simulated images were smoothed to a common resolution of $0.12\arcsec \times 0.12\arcsec$ to erase potential effects of differences in beam position angles.
Deprojected radial profiles of the convolved model images were derived following the approach described in Section \ref{subsec:oui}. These radial profiles were computed in the image plane rather than reconstructed from the visibilities to avoid the introduction of artifacts possibly altering our analysis \citep{Viscardi+2025}.
At each radius, the agreement between the data were calculated using the likelihood $\mathcal{L}\propto\exp(-\chi^2_{\text{dust}}/2)$ using 
\begin{equation}
\label{eq:chi2_dust}
\chi^2_{\text{dust}} = \sum_{j} \left( \frac{I_{\nu,j} - I_{\nu,\text{mod},j}}{\hat{\sigma}_{I_{\nu}, j}} \right)^2 \, ,
\end{equation}
where $j$ is the radial bin, $I_{\nu,j}$ the data intensity in that bin, $I_{\nu,\text{mod},j}$ the model intensity at the same location, and $\hat{\sigma}_{I_{\nu}, j}$ the error on the intensity. The latter was computed as follows: 
\begin{equation}
\label{eq:rad_error_withcalib}
\hat{\sigma}_{I_{\nu}, j} = \sqrt{ \sigma_{I_{\nu}, j}^2 + \delta_{\nu}^2 I_{\nu, j}^2 } \,,
\end{equation}
with $\delta_{\nu}$ being the calibration error at the frequency $\nu$ and $\sigma_{I_{\nu}, j}$ defined by Equation \ref{eq:rad_error}. The calibration errors were taken as $5\%$ for the $1.3$~mm data and the $3.0$~mm data, and as $10\%$ for the $8.9$~mm data. 
The fit was performed between $25$~au and $51$~au, which are the limits of the disc model. The fit calculated the likelihood of each {\sc MCFOST} simulation independently at each radius. Assuming the simulation grid as uniform priors, the posterior distribution of a set of parameters $\Theta$ in a radial bin $j$ was computed as $P(I_{\nu,j}\vert\Theta) = \exp(-\chi^2_{\text{dust}}/2)$. 
As a function of the input $M_{\text{dust}}$, {\sc MCFOST} outputs the surface density and the temperature profiles, respectively noted $\Sigma_d(r)$ and $T_d(r)$. By retrieving these values from the models, our analysis allowed to explore the radial distribution of the dust surface density and temperature. \mdy{Because the underlying models differed in their surface densities, which was set by their total dust masses, the reconstructed surface density profile can deviate from the $p=-1$ power-law chosen for the individual models.}
In each radial bin, the dust temperature was azimuthally and vertically mass-averaged.
The set of parameters is then $\Theta=(\Sigma_d,T_d,a_{max},q)$. Given the posterior distribution, the expected values of the individual parameters in a radial bin $j$ were computed with the following formula:

\begin{equation}
\label{eq:expectation}
E_j(X) = \frac{\sum_i X_i P(X_i \vert I_{\nu,j})}{\sum_i P(X_i \vert I_{\nu,j})} \,,
\end{equation}
with $\left\{X_i\right\}_i$ the set of values taken by the parameter~${X\in\left\{\Sigma_d,T_d,a_{max},q\right\}}$.

\begin{figure*}
\centering
\begin{center}
    \includegraphics[width=\textwidth, trim={0cm 0cm 0cm 0cm},clip]{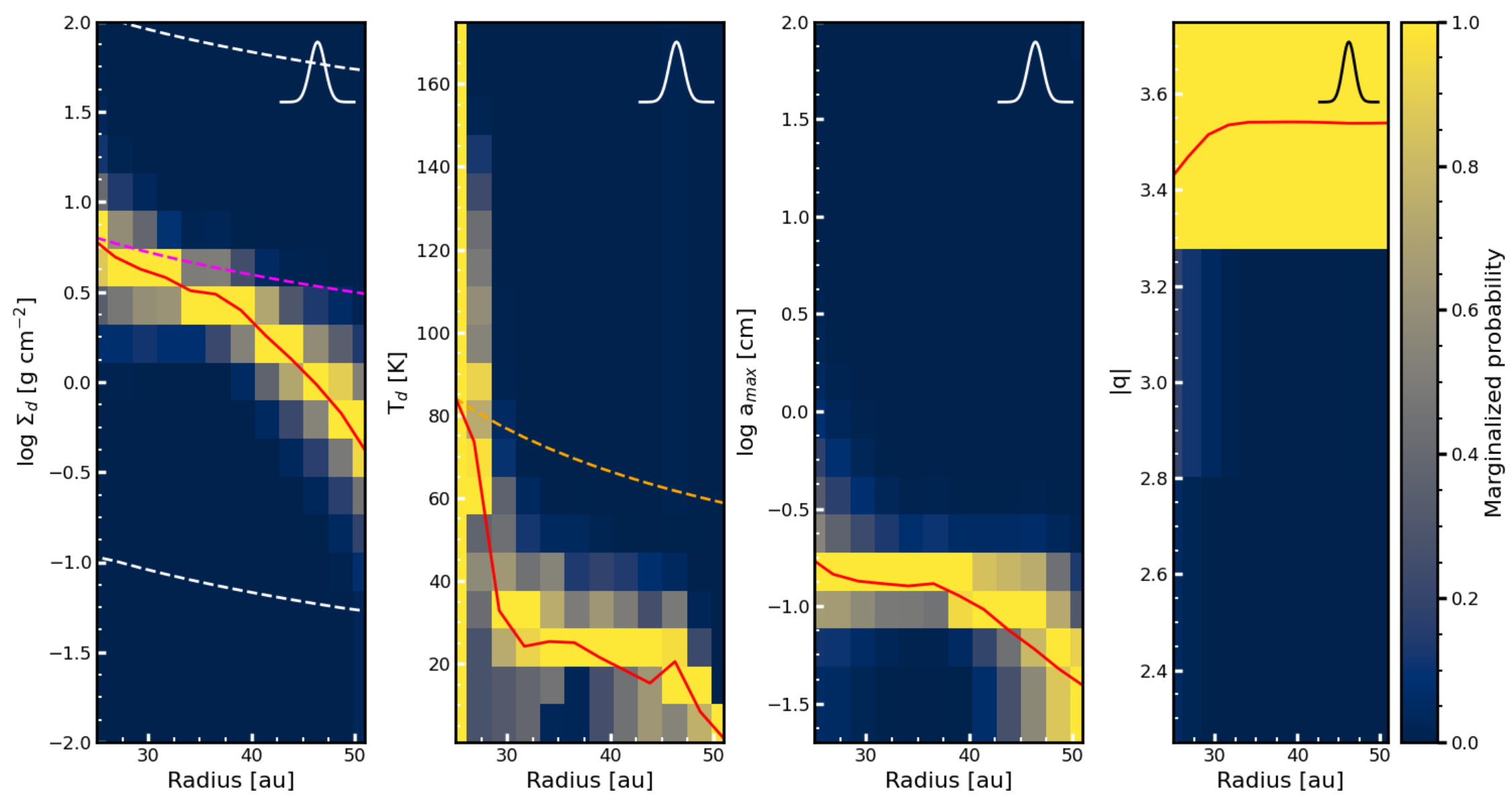}     
    \caption{Marginalized probability distribution of the parameters sampled in Section \ref{subsec:dust_modelling} as a function of radius in the ring of V892~Tau. The probabilities are normalized to the maximum marginalized probability at each radius to ease the visualisation. The red curve shows the expected value for each parameter, computed following Equation \ref{eq:expectation}. The pink dashed line on the surface density panel corresponds to a disc density profile respecting $\Sigma\propto r^{-1}$, while the white dashed lines delimit the simulated parameter space. The orange dashed curve on the temperature panel corresponds to the temperature profile of a passive disc irradiated by a central luminosity of L$_* = 144$ L$_{\odot}$ and  having a flaring angle of $\varphi=0.018$ \citep{ChiangGoldreich1997,Dullemond+2001}. The inset at the top right of each panel corresponds to the geometric mean of the beam size used to smooth the models.}
    \label{fig:posteriors_dust}
\end{center}
\end{figure*}
\begin{figure*}
\centering
\begin{center}
    \includegraphics[width=\textwidth, trim={0cm 0cm 0cm 0cm},clip]{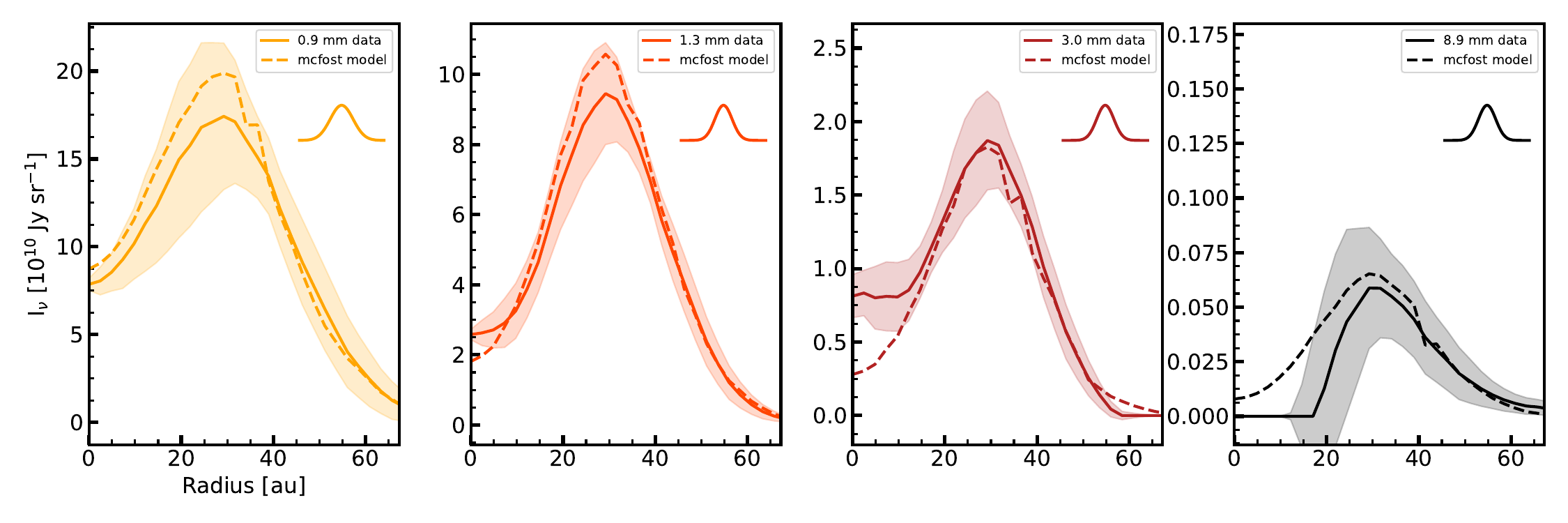}     
    \caption{Deprojected smoothed radial intensity profiles of the data and the best-fit model at each wavelength. The data is indicated by plain lines whilst the best-fit {\sc MCFOST} model at each radius is depicted in dashed lines. The shaded area corresponds to the error on the data radial profiles computed following Equation \ref{eq:rad_error}. From left to right, the panels show the radial profiles at $0.9$~mm, $1.3$~mm, $3.0$~mm, and $8.9$~mm. The $0.9$~mm data is shown in its original resolution and was not used in the modelling procedure. The other observations were smoothed to a common resolution of $0.12\arcsec \times 0.12\arcsec$. The geometric mean of the full beam size is represented by the inset curve at the top right of each panel.
    }
    \label{fig:I_bestfit}
\end{center}
\end{figure*}

The results of the multi-wavelength modelling are showed on Figures \ref{fig:posteriors_dust} and \ref{fig:I_bestfit}.
Figure \ref{fig:posteriors_dust} shows the radial distribution of the dust properties that best reproduce the data, namely the surface density $\Sigma_d$, the mass-averaged dust temperature $T_d$, the maximum grain size $a_{max}$, and the exponent of the grain size distribution $q$. The distribution is shown marginalized to each parameter and normalized to the maximum probability in each radial bin. The red curve corresponds to the expected value of each parameter, calculated following Equation \ref{eq:expectation} as a function of radius. The most likely models show a decrease in surface density from approximately $6.3$~g~cm$^{-2}$ at the cavity edge to approximately $0.4$~g~cm$^{-2}$ in the outer dust ring. This decrease is steeper than a $\Sigma_d\propto r^{-1}$ profile, shown by the pink curve on Figure \ref{fig:posteriors_dust}. \mdy{Assuming equal surface density at the inner disc edge, our modelling shows $\Sigma_d$ to be ten times lower at the outer disc edge compared to a $\Sigma_d\propto r^{-1}$ profile. }

Close the cavity edge, the dust temperature is best modelled by a temperature of $T_d\sim80$~K. Beyond $30$~au, the dust temperature decreases from $\sim30$~K to $\lesssim10$~K. Given the angular resolution of the observations shown by the inset Gaussian curve at the top right of each panel, this decreasing trend is physical. The orange curve on the temperature panel corresponds to a model of a passive disc irradiated by L$_* = 144$~L$_{\odot}$ and flared with an angle $\varphi=0.018$ \citep{ChiangGoldreich1997,Dullemond+2001}. The dust temperature is found lower than the prediction of the passive disc model, implying an efficient shielding of the dust midplane. This efficient shielding could be created by a puffed-up inner rim protecting a flat dust disc, or by dense upper layers \citep{Dullemond+2002,Dong2015}. 
The distribution of $a_{max}$ is flat up to $40$~au, with a value of approximately $0.2$~cm. The observations resolve a decrease in $a_{max}$ beyond $40$~au. The expectation curve adopts the same qualitative behaviour and shows that the largest grains stand in the inner half of the ring. 
Finally, a value of $q=-3.5$ best matches the data at all radii. Beyond $33$ au, the expected $q$ is in agreement with this value. 

For each radial bin, there is one {\sc MCFOST} simulation which best fits the data. The optimal {\sc MCFOST} simulation may vary from one radial bin to another. Hereafter we call \textit{best-fit model} the compilation of parameters producing the best-fitting {\sc MCFOST} models at each radius.
Figure \ref{fig:I_bestfit} shows the deprojected intensity radial profiles of the best-fit model in comparison with the smoothed data. 
Overall, the intensity from the best-fit model shows a good agreement with the data intensity: over the extent of the disc model, \textit{i.e.} from $R_{in}=25$ au to $R_{out}=51$ au, the model intensity is within the error-bars of the data. The best-fit models at $R_{in}$ and $R_{out}$ were assumed as the best-fit models inside $R_{in}$ and outside $R_{out}$, respectively. Beyond $R_{out}$, the best-fit model matches the data closely except at $3.0$~mm due to negatives after $56$~au. Our {\sc MCFOST} models did not include a central emission source, which could explain the mismatch with the data inside $R_{in}$ at $1.3$~mm and $3.0$~mm. At $8.9$~mm, the mismatch comes from the removal of the binary on the data image, which put the data intensity to $0$ inside $\sim20$ au. Beam convolution could spread signal from the dust ring to the innermost regions, but this does not affect our conclusions (see Appendix \ref{app : central_emission}).

\subsection{Mass reservoir}
\label{subsec:dust_mass}

Assuming optically thin emission, the dust mass in the disc can be estimated following the expression \citep{Hildebrand+1983}:
\begin{equation}
\label{eq:dustmass_thin}
    M_{\text{dust}} = \frac{F_{\nu}\; d^2}{\kappa_{abs}\;B_{\nu}(T_d)\;\cos(i)}\,,
\end{equation}
with $M_{\text{dust}}$ being the dust mass in the disc, $F_{\nu}$ the flux at the frequency $\nu$, $\kappa_{abs}$ the absorption opacity at the corresponding frequency, $B_{\nu}(T_d)$ the blackbody intensity at a temperature $T_d$, $i$ the inclination, and $d$ the distance to the system. 
For these computations, the temperature was fixed to $30$~K in agreement with our modelling, the distance to $134.5$~pc \citep{GaiaDR3}, and the fluxes of the dust disc to the $F_{out}$ values listed in Table \ref{table:fluxes}.
For the opacities, we assumed the DSHARP composition and a power-law index of the grain size distribution following our best-fit value of $q=-3.5$. We computed a first set of dust masses by assuming the same opacities than in Section \ref{subsec:dust_modelling}, \textit{i.e.} opacities that were averaged over the dust size distribution up to $a_{max}$. According to our best-fit results, we chose here $a_{max}=0.15$~cm. 
To facilitate comparison with previous studies, we computed a second set of masses using power-law opacities $\kappa_{abs}\propto\nu\,^{\beta}$ normalised to $2.3$~cm$^2$~g$^{-1}$ at $1.3$~mm \citep[e.g.][]{Andrews+2013}. We assumed $\beta=\alpha_{\text{out}}-2$, namely $\beta=0.10$ at $0.9$~mm, $1.3$~mm, and $3.0$~mm, and $\beta=0.78$ at $8.0$~mm and $9.8$~mm. 
The uncertainties on the dust masses were propagated from the flux errors of the outer disc, which are summarised in Table \ref{table:fluxes}. 
Finally, we computed the dust mass from our multi-wavelength analysis by integrating the best-fit surface density profile over the disc extent, with the uncertainties propagated from the $1\sigma$ confidence interval around the best-fit surface densities. All the dust masses are listed in Table \ref{table:masses}.

\begin{table}
    \centering
    \caption{Mass of the dust disc in V892~Tau derived at each wavelength using the methods of Section \ref{subsec:dust_mass}.}
 \begin{tabular}{c c c} 
Wavelength & \multicolumn{2}{c}{M$_{\text{dust}}$}\\ [0.5ex] 
(mm) & \multicolumn{2}{c}{($10^{-4}$ M$_{\odot}$)} \\
 \hline
 & $\kappa_{abs}\propto\nu\,^{\beta}$\,\,\,\tablefootmark{1} & $\kappa_{abs} = \kappa_{\text{DSHARP}}$\,\,\,\tablefootmark{2} \\ 
 \hline \hline
$9.8$ & $6.9 \pm 1.3$ & $220 \pm 40$ \\
$8.0$ & $6.8 \pm 1.1$ & $130 \pm 20$ \\
$3.0$ & $4.6 \pm 0.2$ & $28 \pm 1$ \\
$1.3$ & $4.8 \pm 0.2$ & $6.6 \pm 0.3$ \\
$0.9$ & $5.6 \pm 0.6$ & $4.6 \pm 0.5$ \\
\hline
SED & / & $29^{+17}_{-15}$ \\
\end{tabular}
\tablefoot{At each wavelength, the dust mass was derived from the values of $F_{out}$ listed in table \ref{table:fluxes} using Equation \ref{eq:dustmass_thin}. The last row indicates the dust mass derived from our best-fit model described in Section \ref{subsec:dust_modelling}.
\tablefootmark{1}{This column corresponds to dust masses calculated assuming power-law opacities.} \\
\tablefootmark{2}{This column corresponds to dust masses calculated assuming DSHARP opacities averaged over the dust size distribution.}
}
\label{table:masses}
\end{table}

Our best-fit model has a dust mass of $29_{-15}^{+17} \times 10^{-4}$~M$_{\odot}$. Using Equation \ref{eq:dustmass_thin} and power-law opacities, the dust mass of the disc is comprised between $4 \times 10^{-4}$~M$_{\odot}$ and $7 \times 10^{-4}$~M$_{\odot}$. Using Equation \ref{eq:dustmass_thin} and DSHARP opacities, the dust masses are also lower than the SED-based value up to $3.0$~mm. It suggests than the emission below $3.0$~mm is significantly optically thick.
We show in the next Section \ref{subsec:opt_depth} that our best-fit model is indeed optically thick at $3.0$~mm and below. Dust mass estimates derived from SED analysis have been shown to be more robust and reliable than those based on flux measurements, which tend to underestimate the dust mass \citep{BalleringEisner2019,Ribas+2020,Viscardi+2025}. At larger wavelengths, the flux-based dust masses derived using DSHARP opacities exceed those obtained from SED-based estimates. We discuss the reasons of this mismatch in the context of other studies in Section \ref{subsec:context_discussion}.

\subsection{Optical depth}
\label{subsec:opt_depth}

\begin{figure}
\centering
\begin{center}
    \includegraphics[width=\columnwidth, trim={0cm 0cm 0cm 0cm},clip]{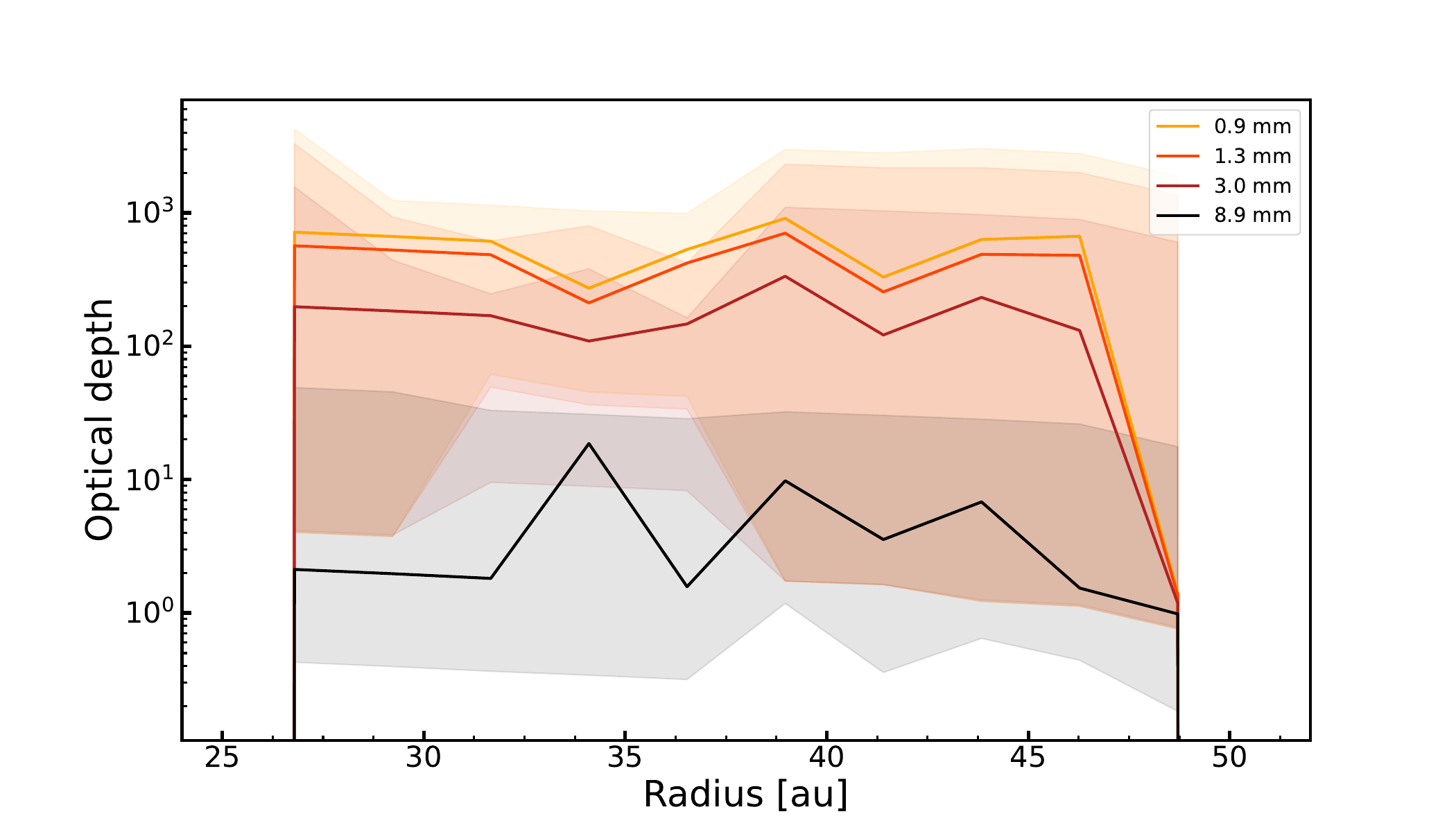}     
    \caption{Optical depth radial profiles corresponding to the best-fit model. The plain curves in yellow, orange, dark red, and black correspond to the optical depth of the model at $0.9$~mm, $1.3$~mm, $3.0$~mm, and $8.9$~mm, respectively. The shaded areas indicate the errors computed as the $1\sigma$ confidence interval.}
    \label{fig:tau_profiles}
\end{center}
\end{figure}

Using a Monte-Carlo method, {\sc MCFOST} is able to compute the optical depth in the line of sight \citep{Pinte+2006}.
The optical depth radial profiles $\tau(r)$ corresponding to best-fit model can be found on Figure \ref{fig:tau_profiles} for each wavelength. As with the geometrical fitting, uncertainties were defined as the bounds of the $1\sigma$ confidence interval around the best-fit values \citep{Andrae2010}. 
The best-fit model is found optically thick at all radii within the uncertainties from $0.9$~mm to $3.0$~mm. At $8.9$~mm, the best-fit model has $\tau>1$ but the error-bars show that models with $\tau<1$ are also compatible with the data within $1\sigma$. 
Having optically thick emission between $0.9$~mm and $3.0$~mm and optically thinner emission between $3.0$~mm to $8.9$~mm is consistent with the spectral indices calculated in Section \ref{subsec:sp_ind}.

\subsection{Fragmentation velocity of dust grains}
\label{subsec:vfrag}

When two dust grains collide, the outcome of the collision depends on the relative velocity, the composition, and the geometry of the grains \citep{Birnstiel2024}. At low relative velocities, grains may stick together. At higher velocities, they may bounce off each other or fragment. There is a critical velocity, noted $v_{\text{frag}}$, above which collisions lead to fragmentation rather than growth. In this Section, we estimate this critical velocity using the expected profiles $\Sigma_d(r)$, $T_d(r)$, and $a_{max}(r)$ shown by the red curve on Figure \ref{fig:posteriors_dust}.
The Stokes number, noted $\text{St}$, describes the aerodynamical coupling between gas and dust grains. We assumed that dust and gas are perfectly coupled, \textit{i.e.} $\text{St}\ll1$. From Equation~$4$ in \cite{Rosotti+2020}, we obtained $\text{St}=0.07\pm0.04$ using $\Sigma_d(r=30\,\text{au})=30\pm10$~g~cm$^{-2}$ and $a_{max}(r=30\,\text{au})=0.15\pm0.08$~cm, which justifies our assumption.

\begin{figure}
\centering
\begin{center}
    \includegraphics[width=\columnwidth, trim={0cm 0cm 0cm 0cm},clip]{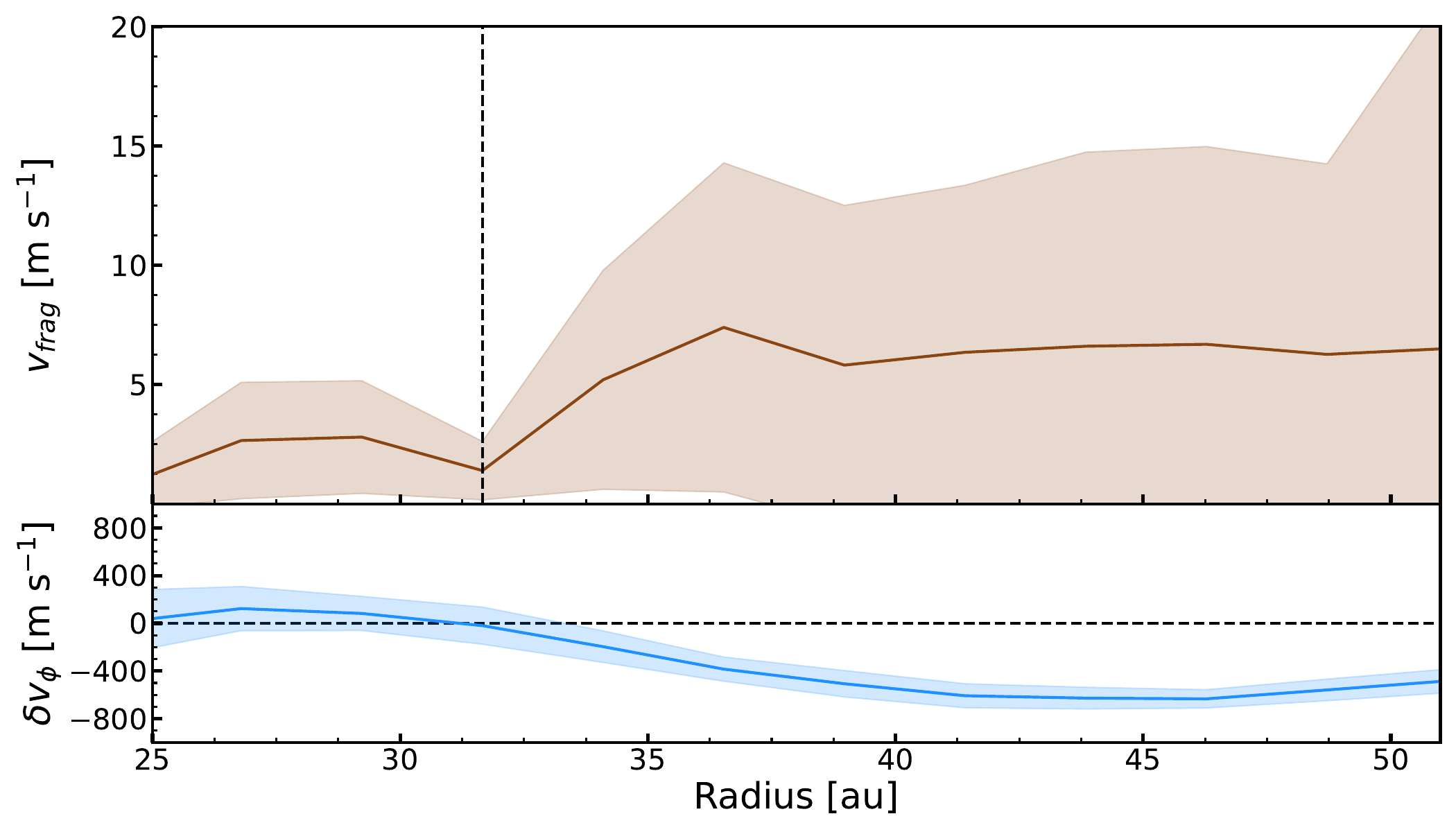}     
    \caption{Fragmentation velocity of dust grains (top) calculated from Equation \ref{eq:vfrag} and deviation from Keplerian rotation speed from \cite{Alaguero+2024} (bottom) as a function of radius in the ring. The shaded area corresponds to the $1\sigma$ error. The vertical dashed line on the top panel corresponds to the radius where temperature profile passes below $25$ K. The horizontal dashed line on the bottom panel correspond to the Keplerian velocity.}
    \label{fig:vfrag}
\end{center}
\end{figure}

The fragmentation velocity $v_{\text{frag}}$ can be defined as \citep{Zagaria+2023}:
\begin{equation}
\label{eq:vfrag}
v_{\text{frag}} = 2.85 \, \frac{\pi}{2} \, \rho_s \, a_{max} \, \frac{c_s}{\Sigma_{g}} \, \left( \frac{\alpha_{\text{turb}}}{\text{St}} \right)^{1/2} \, ,
\end{equation}
with $\rho_s$ being the intrinsic grain density, $a_{max}$ the maximum grain size assumed limited by fragmentation, $c_s$ the sound speed, $\Sigma_{g}$ the gas surface density, and $\alpha_{\text{turb}}$ a parametrisation of the turbulence in the disc \citep{OrmelCuzzi2007}. The previous quantities are calculated as follows \citep{Rosotti+2020}:
\begin{eqnarray}
\label{eq:turb_stokes}
\Sigma_{g} &=& \delta_{gd} \, \Sigma_{d} \, , \\
c_s &=& \sqrt{\frac{k_b T_g}{\mu m_p}} \, , \\
\frac{\alpha_{\text{turb}}}{\text{St}} &=& 2 \, \Sigma_{d} \, \frac{v^2_{k}}{c^2_{s}}  \, \frac{\delta v_{\phi}}{v_{k}} \, \left(\frac{\text{d} \Sigma_{d}}{\text{d}r}\right)^{-1} \,,
\end{eqnarray}
with $\delta_{gd}$ being the gas to dust mass ratio, $k_b$ the Boltzmann constant, $T_g$ the gas temperature, $\mu$ the mean molecular weight, $m_p$ the proton mass, $v_k$ the Keplerian speed, and $\delta v_{\phi}$ the deviation of the azimuthal rotation speed from the Keplerian speed. 
We assumed $\delta_{gd}=100$ and $\mu=2.33$ in atomic units. We also assumed that gas and dust were at local thermal equilibrium, so that $T_g=T_d$. The DSHARP dust grain composition had a grain density of $1.68$ g cm$^{-3}$. Finally, $\delta v_{\phi}$ was estimated from the rotation curve of the $^{12}$CO emission using {\sc Discminer} \citep{Izquierdo+2021,Izquierdo+2023} in \cite{Alaguero+2024} (see their Appendix~C). 
Given our assumptions, Figure \ref{fig:vfrag} shows the radial profiles of $v_{\text{frag}}$ and $\delta v_{\phi}$.

From the cavity edge at $25$~au to approximately $32$ au, the fragmentation velocity profile is flat and remains below $5$~m~s$^{-1}$. After that, the fragmentation velocity rapidly increases to reach $\sim8$~m~s$^{-1}$ at $\sim33$ au. The fragmentation velocity then remains roughly constant up to the outer radius of the ring.
Although $v_{\text{frag}}$ is poorly constrained in the outer parts of the ring, the ring seems divided in two distinct parts. Interestingly, the dust temperature drops below $25$~K at a radius in between the low and high $v_{\text{frag}}$ regions of the disc, as indicated by the vertical dashed line on Figure \ref{fig:vfrag}. This temperature lies within the expected freeze-out range for CO molecules onto dust grains in the disc midplane, typically $20-30$~K \citep[e.g][]{Pinte+2018b,ObergWordsworth2019}. Inside the snowline, dust grains are bare and often represented by silicate. Outside the snowline, dust grains are coated by frozen volatile material, which affects their fragmentation velocity. Whilst the precise fragmentation velocity of grains depends from one study to the other, bare grains are generally found to have lower fragmentation velocities than icy grains \citep{Drazkowska+2023}. Our results then suggest a spatial segregation in the grain composition, which appears to coincide with the position of the CO snowline in the disc.

\section{Discussion}
\label{sec:discussion}

\subsection{Dust dynamics and properties in V892~Tau}
\label{subsec:discussion_vtau}

The dust thermal emission shows a ring around the central binary of V892~Tau. The inner radius of the ring is found at approximately $20$~au in the observations and at $25$~au by our modelling. We found that the greatest dust surface densities, dust temperature, and dust maximum grain size are achieved at the inner rim of the disc (see Figure \ref{fig:posteriors_dust}). 
A dust cavity cavity of $25$ au would represent $3.7$~times the semi-major axis of the central binary \citep{Vides+2023}, which is in good agreement with tidal truncation models \citep{MirandaLai2015,Ragusa+2020}. At the edge of this tidally carved cavity, the central binary star would naturally create a pressure bump halting the radial drift of dust particles \citep{Chachan+2019,Coleman+2022}. Dust particles can then pile-up at the cavity edge, enhancing the local dust to gas mass ratio up to one order of magnitude \citep{Poblete+2019}.

It is unclear to what extent local concentrations of dust would effectively promote grain growth or not, especially in multiple stellar systems. Close binaries are known to perturb the disc kinematics and to excite high collisional velocities between dust grains, inhibiting their growth \citep{MarzariThebault2019}. \cite{Eriksson+2025} found that relative velocities of dust grains in spiral arms induced by planets were always larger than $5$~m~s$^{-1}$. Extrapolating to stellar companions close to the cavity edge of circumbinary discs, the relative velocities of dust grains would exceed the fragmentation velocities of $\sim8$~m~s$^{-1}$ found in Section \ref{subsec:vfrag}. On top of that, the value of $q=-3.5$ obtained in our modelling is consistent with dust size distributions produced by collisional cascades in which the size of dust grains is limited by fragmentation \citep{Birnstiel+2012}.
Our analysis then suggests that dust growth in V892~Tau is limited by the perturbations of the central binary. A detailed modelling of grain growth in circumbinary discs will be addressed in a following study, helping to conclude more robustly on this matter.

\subsection{V892~Tau in the context of other multi-wavelength studies}
\label{subsec:context_discussion}

We performed a multi-wavelength analysis of the mm emission in V892~Tau to constrain its dust properties. In this Section, we place our results in the context of recent multi-wavelength studies of other systems. 
The validity of such a comparison relies on a common assumption for the grain opacities, as different grain opacities would naturally yield in different dust surface density, temperature, and grain size \citep[e.g.][]{Sierra+2025}. As a consequence, we compare this work with those of other studies that likewise employ the DSHARP opacity prescription. Whilst we used a 3D full radiative transfer method, the analysis of multi-wavelength observations is generally done based on 1D analytical radiative transfer of an isothermal dust slab. We compare these two methods V892~Tau in Appendix using the practical case of V892~Tau \ref{app : dust_slab}.  

Works using dust slab models have consistently shown nearly flat maximum grain size profiles over the extent of dust rings \citep{Sierra+2021,Guidi+2022,Jiang+2024}, a feature that we retrieve for V892~Tau at radii below $40$ au. Smooth $a_{max}$ could indicate grain sizes limited by fragmentation \citep{Jiang+2024}. In that case, our modelling suggests fragmentation velocities of the order of $\lesssim8$~m~s$^{-1}$ (see Figure \ref{fig:vfrag}). 
Multi-wavelength studies commonly found $a_{max}$ to peak around $1$~cm \citep[e.g.][]{Sierra+2024}, which is higher that the measurement of $0.2$~cm that we obtain for V892~Tau. Beyond indicating that dust growth seems limited, lower maximum grain sizes result in higher grain opacities for a given amount of mass (see Appendix \ref{app : opacities}), in turn resulting in higher optical depths. We indeed found optical depths at least $10$ times larger than what found by \cite{Sierra+2024} for various systems. 
We note that larger dust surface densities could also lead to larger optical depths. The surface density of V892~Tau sits in the $10^{-1}-10^1$~g~cm$^{-2}$ range commonly found in discs \citep[e.g.][]{Jiang+2024,Sierra+2024}. Integrating this surface density gives a mass of approximately $930$ M$_{\oplus}$, which places V892~Tau on the high-mass end of protoplanetary discs in the Taurus star-forming region and above the minimum-mass solar nebula \citep{Weidenschilling1977,BalleringEisner2019,Ribas+2020}. These dust mass measurements need once again to be contextualized given the assumptions done on the grain opacities. Based on a sample of $21$ sources, \cite{Garufi+2025} found that the DSHARP opacities resulted in the highest dust mass estimates at $1$~cm compared to other commonly assumed grain opacities. This effect stems from the steep opacity spectral index $\beta$ of the DSHARP opacities \citep{Birnstiel+2018}, which naturally yields higher dust mass estimates at cm wavelengths. While this may account for the significantly higher flux-based dust masses of V892~Tau compared to the SED-based estimate at $8.0$~mm and beyond, \cite{Viscardi+2025} demonstrated that flux-based dust masses are generally underestimated relative to those derived from SED modelling. This questions the suitability of the DSHARP opacities for representing grain properties in V892~Tau at wavelengths of $\geq8.0$~mm. Further support for this comes from the inability of dust slab models to reproduce the observed intensity at $8.9$~mm when using the DSHARP opacities (see Appendix \ref{app : dust_slab}).

\subsection{Limitations and caveats}
\label{subsec:limits}

This work is based on full 3D radiative transfer calculations which go beyond the 1D analytical model commonly used to study dust properties in resolved discs \citep[e.g.][]{Riviere-Marichalar+2024}. Our results need to be considered given the assumptions made on the grain opacities. Different opacity prescriptions would influence the modelled dust properties and emission profiles. Except in the case of lower absorption opacities which would make $a_{max}$ difficult to constrain, variations in grain composition are not expected to affect the qualitative trends in dust properties. As the DSHARP opacities employed in this study are among the lowest at wavelengths $\gtrsim1$~mm \citep{Birnstiel+2018}, we argue that this condition is satisfied in the present case. In our modelling, we also excluded solutions with $a_{max}<0.03$~cm to avoid degeneracies in opacity. These potential solutions would have resulted in significantly higher $\Sigma_d$ and cooler $T_d$ \citep[e.g.][]{Macias+2021}. \mdy{In case of optically thick emission and fixed $\Sigma_d$, solutions with $a_{max}<0.03$~cm would imply higher $T_d$ than found in our work \citep{Chung+2024}. However, if we fix $T_d$ to Equation \ref{eq:prior_T}, and fit an optically thick 1D dust slab model (see Appendix \ref{app : dust_slab}), we obtain $a_{max}$ from $0.3$~mm to $3$~mm. This result is consistent with the results described in Section \ref{subsec:dust_modelling}, showing that higher $T_d$ would not affect our conclusions in case of optically thick emission at all wavelengths.   } 

Angular resolution limits the precision of our results, as all the models were convolved to the angular resolution of the observations. This convolution may have spread signal in the image, possibly mimicking bright regions that are in reality empty. To address this issue, we fitted the disc extent in the uv plane and then sampled the dust properties in the regions strictly within the fitted disc model.
By convolving the models and performing an independent fit at each radius, we ignored correlations between neighbouring radial bins introduced by the convolution with the beam. In Appendix \ref{app : unique_mcfost_model}, we show that these correlations are negligible by presenting a single radiative transfer model able to reproduce the observations. \mdy{On a related note, while beam dilution can induce errors in the temperature profile, it does not prevent to accurately constrain the maximum grain size at the position of the ring \citep{Viscardi+2025}.}

Our working assumptions for the radiative transfer models could also challenge the validity of our work. \cite{Alaguero+2024} proved the triplicity of V892~Tau whilst our model only accounted for the central binary. Given the M3 stellar type of V892~Tau~NE and its $\sim500$~au separation with the double A5 star at the centre of the ring, the contribution of V892~Tau~NE to the illumination of the circumbinary disc is negligible. Binary stars often break the axisymmetry of discs, which is assumed by our disc model. Radial profiles provide the average intensity at each radius and are useful for studying axisymmetric discs, but V892~Tau shows signs of azimuthal asymmetries, such as a disc wall. These features add uncertainty to the interpretation of the results, especially in the inner regions. Brightness asymmetries resulted in a larger radial intensity scatter, which was translated by larger error-bars in the intensity radial profiles following Equation \ref{eq:rad_error}. 
Finally, it is clear that increasing the sampling of the parameter space would provide a better fit to the data. This would however require extra computational costs, exponentially increasing with the accuracy demanded.







\section{Conclusion}
\label{sec:conclusion}

Building upon the efforts of \cite{Alaguero+2024} in the modelling of V892~Tau, we presented a multi-wavelength study of the circumbinary ring of the system using interferometric observations at $0.9$~mm, $1.3$~mm, $3.0$~mm, $8.0$~mm,  and $9.8$~mm. 
We first analysed the spectral index of the emission to determine its physical origin as a function of space. 
We then performed full 3D radiative transfer simulations to fit the geometry of the disc before further modelling the dust properties. This allowed us to explore the dust surface density, dust temperature, dust maximum grain size, and exponent of the grain size distribution as a function of radius.
Our main findings are summarized as follows:

\begin{enumerate}
    \item The ring is detected and resolved from $0.9$~mm to $9.8$~mm. Emission is also found at the centre of the ring, with an increasing contribution to the total flux with wavelength. From spectral index maps, the central emission is likely related to non-dust processes taking place in the vicinity of the $\sim6$ M$_{\odot}$ inner binary. In the ring, the spectral indices are naturally explained by optically thick dust thermal emission up to $3.0$~mm, which gets optically thinner at longer wavelengths.

    \item Our geometric best-fit is a ring with an inner radius $R_{in}=25^{+4}_{-2}$~au and an outer radius $R_{out}=51\pm3$~au. The disc is found with an average inclination of $i=55.5\degree^{+0.5}_{-0.3}$ and position angle of PA$=52.7\degree^{+0.6}_{-0.4}$. 

    \item Our spectral analysis reveals that the disc inner edge is dense ($\Sigma_d\approx6.3$~g~cm$^{-2}$), hot ($T\approx80$~K), and with a maximum grain size of approximately $0.2$~cm. All of these values decrease with radius. The decrease in surface density is steeper than $\Sigma\propto r^{-1}$, whilst temperature drops below $30$~K at $\sim30$~au. Maximum grain size shows a plateau up to $\sim40$~au. At all radii, the data are best modelled with an exponent of the grain size distribution of $-3.5$. Our best model reproduces the radial intensity profiles within the error-bars at all wavelength from $R_{in}$ to $R_{out}$.

    \item We find that the disc contains a dust mass of $29^{+17}_{-15}\times10^{-4}$ M$_{\odot}$, which is greater than what obtained with flux-based estimates up to $3.0$~mm due to optical depth effects. Our models indeed indicate that the disc is optically thick at all radii up to $3.0$~mm. At larger wavelengths, optical depths down to $0.3$ are compatible with the data at $1\sigma$.  

    \item The disc is divided in two parts when it comes to the fragmentation velocity of dust grains: we found $v_{\text{frag}}\approx5$~m~s$^{-1}$ up to $32$~au, and $v_{\text{frag}}\approx8$~m~s$^{-1}$ beyond that. Interestingly, the location of this separation coincides with the potential location of the CO snowline. We suggest that the relatively small maximum grain size derived in this study, compared to values commonly reported in the literature, may result from tidal perturbations induced by the inner binary.

\end{enumerate}

To date, the dust content of protoplanetary discs has been extensively characterised in single-star systems. To enable meaningful statistical comparisons with discs around multiple stars, additional multi-wavelength studies like this one are essential. This will require more high-resolution observations of multiple stellar systems up to centimetre wavelengths. Such observational efforts are essential for advancing our understanding of the initial conditions of planet formation in multiple stellar environments and for laying the groundwork for a comprehensive theory of planet formation across different stellar configurations.



\begin{acknowledgements}
    The authors thank the anonymous referee for their valuable comments, which have helped improve this work. The authors are sincerely grateful to Feng Long for generously sharing the calibrated and reduced VLA data used in this work. 
    A.A. gratefully acknowledges Francsesco Zagaria for his valuable insights during discussions of this work. A.A. would like to thank Jean-François Gonzalez for his precious advice during the review of this work.
    The authors thank the ARC for having produced the calibrated ALMA datasets used in this work.
    This project has received funding from the European Research Council (ERC) under the European Union Horizon Europe programme (grant agreement No. 101042275, project Stellar-MADE). 
    A.R. has been supported by the UK Science and Technology Facilities Council (STFC) via the consolidated grant ST/W000997/1 and by the European Union’s Horizon 2020 research and innovation programme under the Marie Sklodowska-Curie grant agreement No. 823823 (RISE DUSTBUSTERS project).
    J.M. acknowledges support from FONDECYT de Postdoctorado 2024 \#3240612 and from the Millennium Nucleus on Young Exoplanets and their Moons (YEMS), ANID - Center Code  NCN2024\_001.
    This work makes use of {\sc Numpy} \citep{numpy}, {\sc Matplotlib} \citep{Matplotlib}, and {\sc astropy} \citep{astropy2013,astropy2018,astropy2022}. 
    The data underlying this article will be shared on reasonable request to the corresponding author. The code {\sc MCFOST} used in this work is publicly available at \url{https://github.com/cpinte/mcfost}.
\end{acknowledgements}

\bibliographystyle{aa} 
\bibliography{biblio}

\clearpage

\begin{appendix}

\section{Removing of central emission}
\label{app : free-free}

\begin{figure}
\centering
\begin{center}
    \includegraphics[width=\columnwidth, trim={0cm 0.5cm 4cm 2cm},clip]{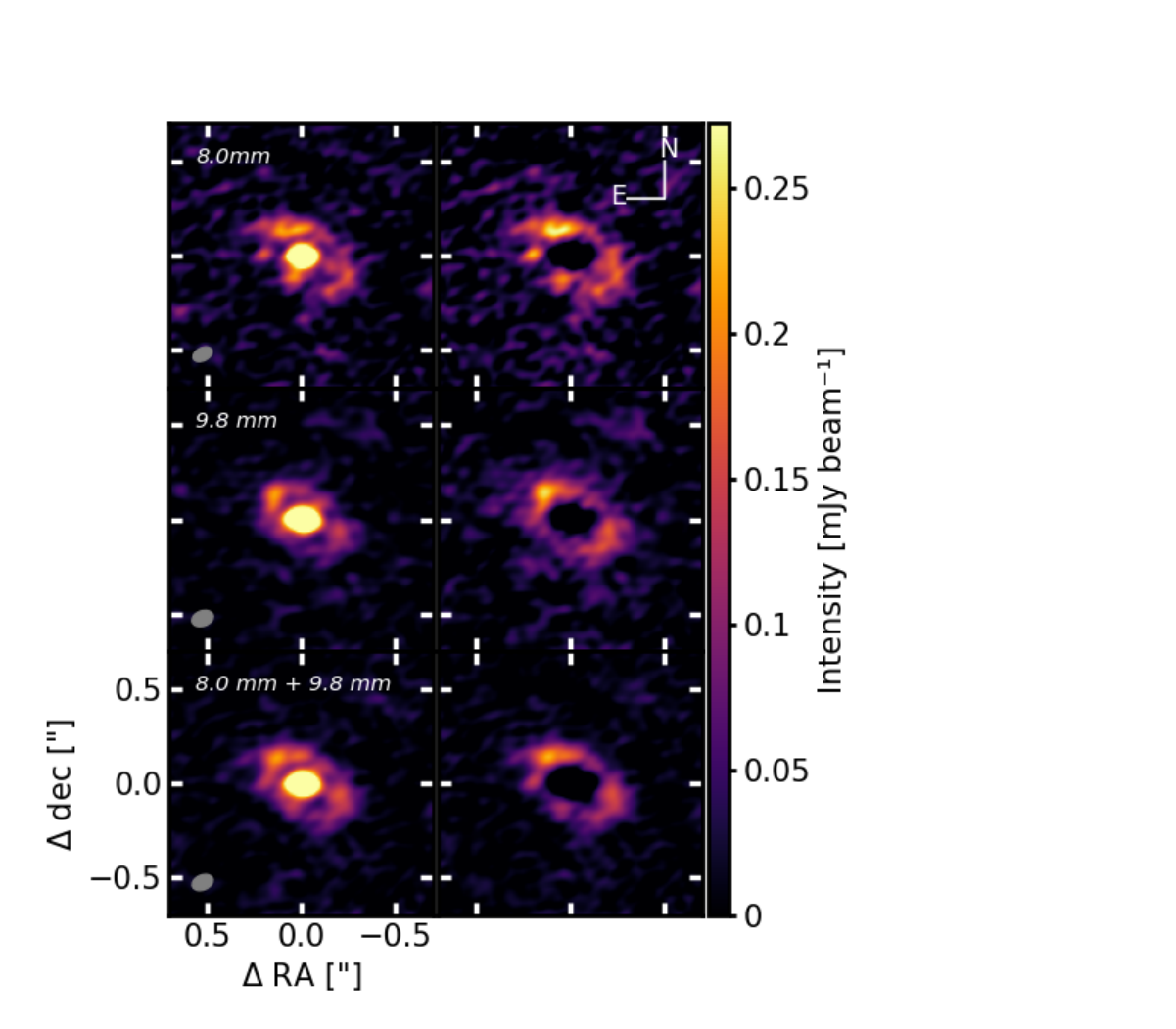}     
    \caption{Continuum maps of the original VLA datasets (left column) and of the VLA datasets where the central binary has been removed (right column). From top to bottom, the images show the data at $8.0$ mm, the data at $9.8$ mm, and the combination of both datasets. The size of the synthesized beam, which is the same for before and after the binary model subtraction, is represented by the grey ellipse at the bottom left of each original image.  Orientation of the sky plane is indicated in the top right panel.}
    \label{fig:binsub}
\end{center}
\end{figure}

We describe in this section the methods used for the removal of the central emission associated with the inner binary of V892~Tau. This was done in an attempt to mitigate the free-free emission contribution to the radial profile. Doing so then ensures that the radial profiles are representative of the dust thermal emission. Images resulting from the process described below can be found in Figure \ref{fig:binsub}.

We fit the data of V892~Tau at $37.5$ GHz and $30.5$ GHz by a model of binary stars consisting in two unresolved components. In the visibility domain, this two point sources model can be defined as follows :

\begin{equation}
V_{mod} = F_1 \exp(-2i\pi \langle\Vec{r_1}, \Vec{f} \rangle) + F_2 \exp(-2i\pi \langle\Vec{r_2}, \Vec{f} \rangle)
\end{equation}

with $\Vec{f}=$($u$,$v$) the coordinates in the uv-plane, $\Vec{r_j}=$($\Delta\alpha_j$, $\Delta\delta_j$) the offsets in right ascension and declination of the star j in the sky plane with respect to the phase centre, and $F_j$ the flux of the star j.

\begin{figure*}[tbp]
\centering
\begin{center}
    \includegraphics[width=\columnwidth, trim={0cm 0cm 0cm 0cm},clip]{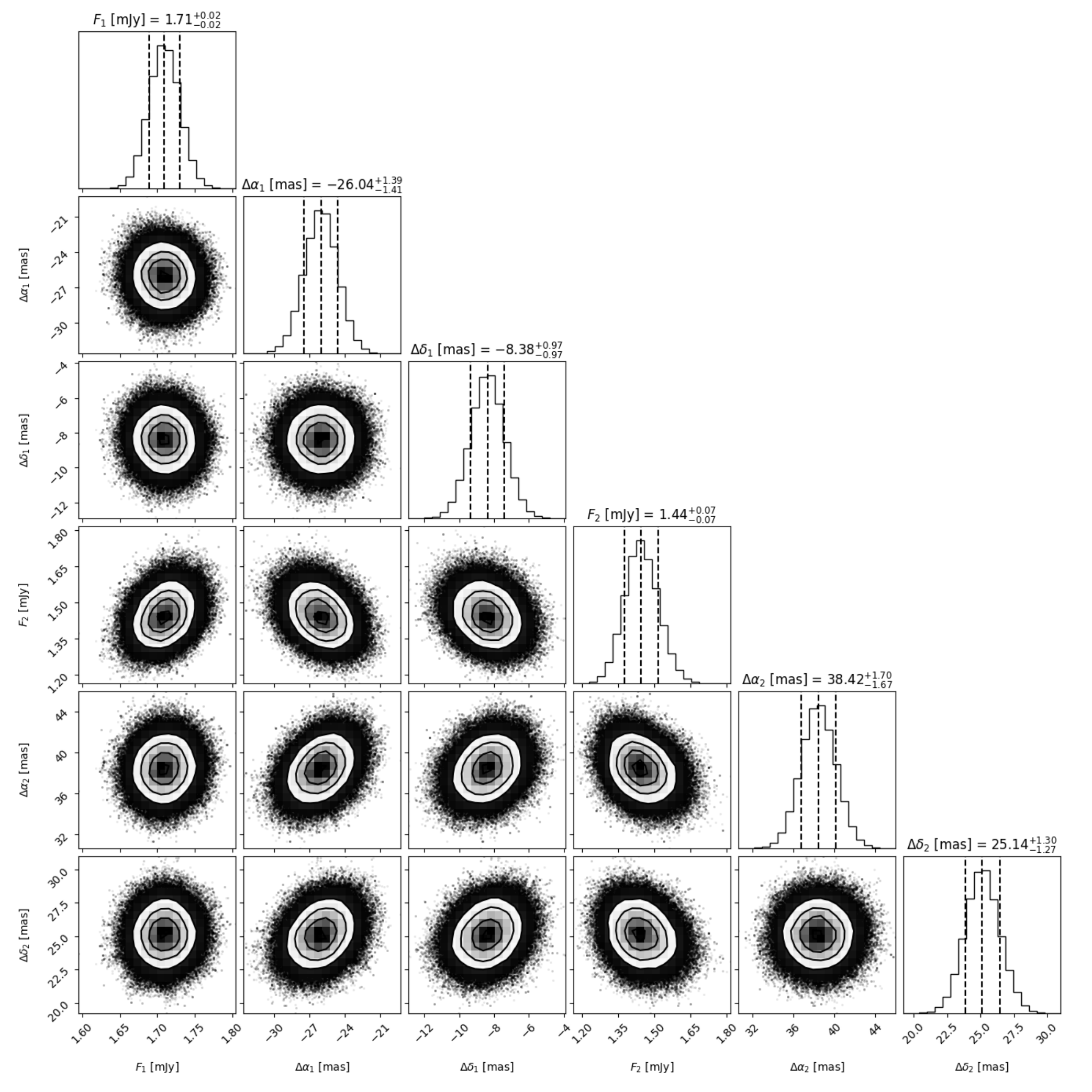}  
    \includegraphics[width=\columnwidth, trim={0cm 0cm 0cm 0cm},clip]{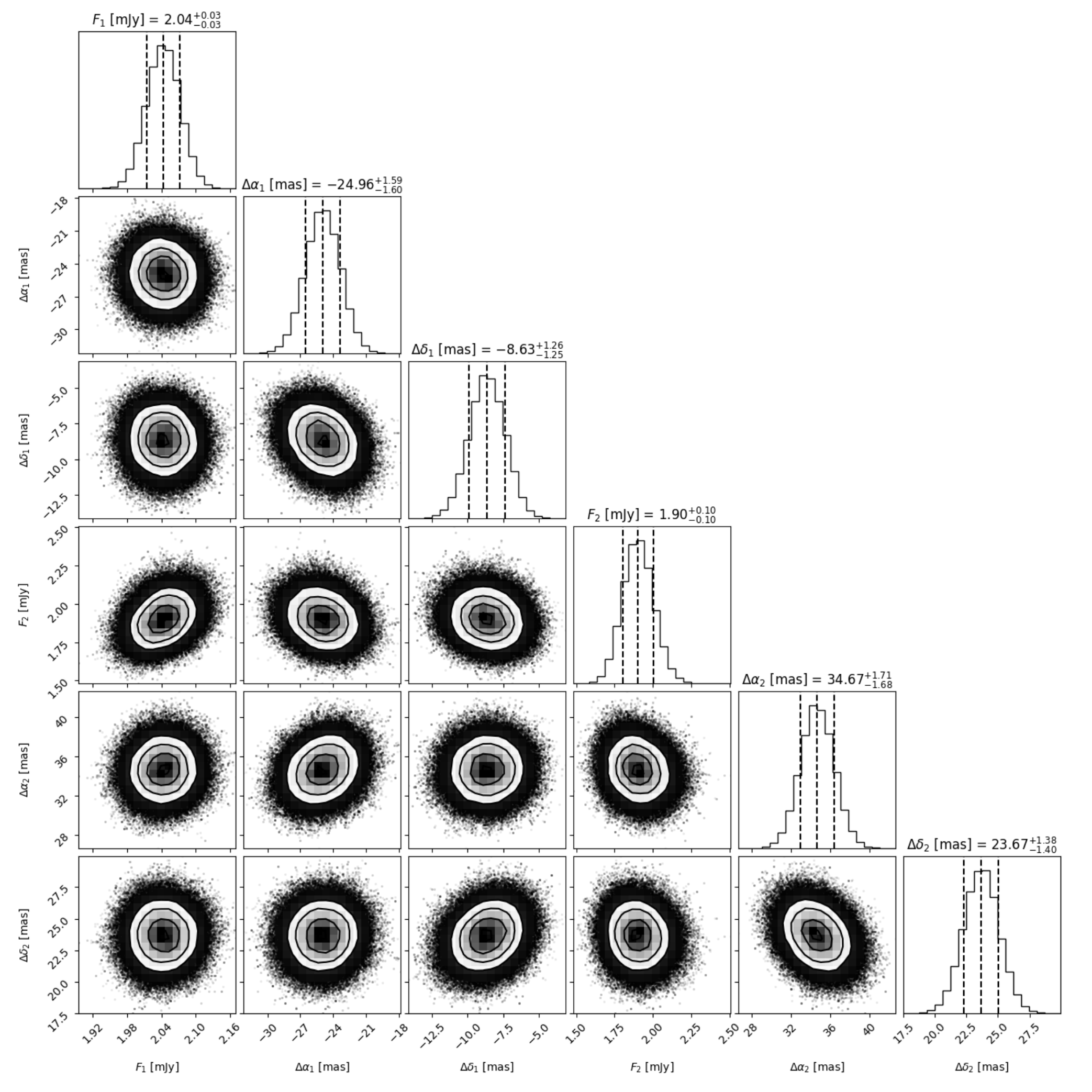} 
    \caption{
    Corner plot of the marginalized posterior distributions after the MCMC sampling for the removal of the binary at $37$ GHz (left) and at $30$ GHz (right). ($\Delta\alpha_j$, $\Delta\delta_j$) are the offsets in right ascension and declination of the star j in the sky plane with respect to the phase centre, and $F_j$ the flux of the star j. The units are written in brackets next to each parameters. For each parameter, best-fit values are indicated and computed as the median values of the posterior distributions. The errors are computed as the $16$th and $84$th percentiles of the posterior distribution.
}

    \label{fig:remove_bin}
\end{center}
\end{figure*}

We fitted this model to the VLA interferometric data at $30.5$ GHz and $37.5$ GHz using the MCMC code {\sc emcee} \citep{emcee} initiating $128$ walkers going for $10000$ steps. We provide to the code a Gaussian likelihood to estimate the quality of the fit, which is defined as follows :
\begin{equation}
 \mathcal{L} \propto \exp( -0.5  \sum_j 37w_j ( \Re(V_j - V_{mod, j})^2 +  \Im(V_j - V_{mod, j})^2 ) )
\end{equation}

with $j$ the index for visibility points, $V$ the data visibilities, $V_{mod}$ the model visibilities and $w$ the weights associated with the data visibilities. As in \cite{Long+2021} (their note 21), we correct the weights by a factor $37$ to represent the standard deviation from the visibility points. The walkers quickly converged to find the positions of the stars in each band. The resulting distribution of walkers is shown in Figure \ref{fig:remove_bin}, after having discarded the first $2000$ steps. We recover different values from \cite{Long+2021} for the best-fit parameters of the binary. This difference can arise because they performed their fitting solely on the VLA A-array data, the most extended where the binary is properly resolved. In our case, the fit is performed on the A-, B-, and C- array combined data, on which the binary is only marginally resolved. However, we stress that these differences are not important in our study, which attempts at removing the central emission without any astrometric considerations. We refer the reader to \cite{Long+2021} for a more precise fitting. 
The best-fit binary model was sampled at the same uv coverage as the observations at $30.5$ GHz and $37.5$ GHz before being subtracted to the visibility data at each of these wavelengths. Finally, the binary-subtracted datasets were imaged together using \textit{tclean} with a Briggs weighting parametrized by a robust parameter of $0.5$ to create the binary-subtracted VLA image used in our spectral analysis.

\FloatBarrier

\section{Spectral index radial profiles}
\label{app : sp_ind_rad}

\begin{figure*}[b]
\centering
\begin{center}
    \includegraphics[width=\textwidth, trim={0cm 0cm 0cm 0cm},clip]{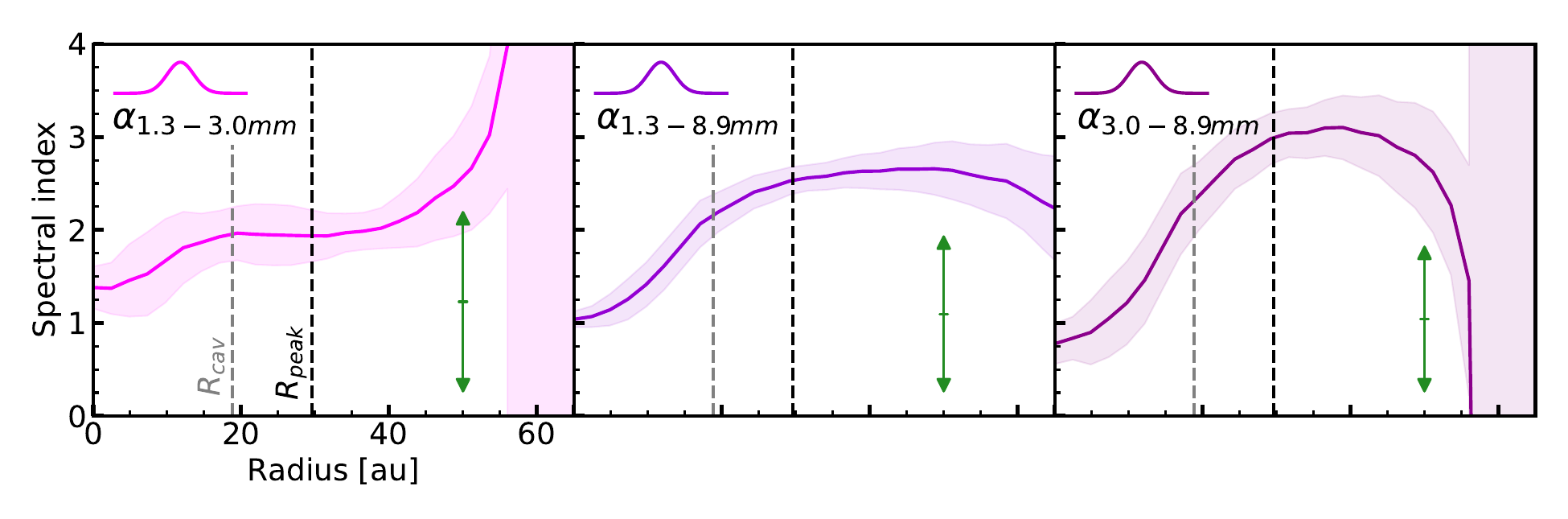}  
    \caption{Spectral index radial profiles calculated following Equation \ref{eq:flux_spectral_index} between $1.3$ mm and $3.0$ mm (left panel), between $1.3$ mm and $8.9$ mm (middle panel), and between $3.0$ mm and $8.9$ mm (right panel). The shaded area correspond to the error, computed from the $1\sigma$ standard deviation of the intensity at each radius. The Gaussian inset indicates the geometric average of the beam size, knowing that all the observations have been convolved to a common resolution of $0.12\arcsec\times0.12\arcsec$. The green arrow in each panel represents the maximum shift possibly introduced by flux calibration systematic errors, which applies equally at each radius. Grey and black vertical dashed lines correspond respectively to the cavity radius $R_{\text{cav}}$ and peak intensity radius $R_{\text{peak}}$ at $1.3$ mm. Beyond $56$ au, negative values in the $3.0$ mm data make the calculations fail.
}
    \label{fig:spectral_index_rad}
\end{center}
\end{figure*}

The spectral index radial profiles can be found in Figure \ref{fig:spectral_index_rad}.

\section{Opacities}
\label{app : opacities}

\begin{figure*}
\begin{minipage}{\textwidth}  
    \centering
    \includegraphics[width=\textwidth, trim={0cm 0cm 0cm 0cm},clip]{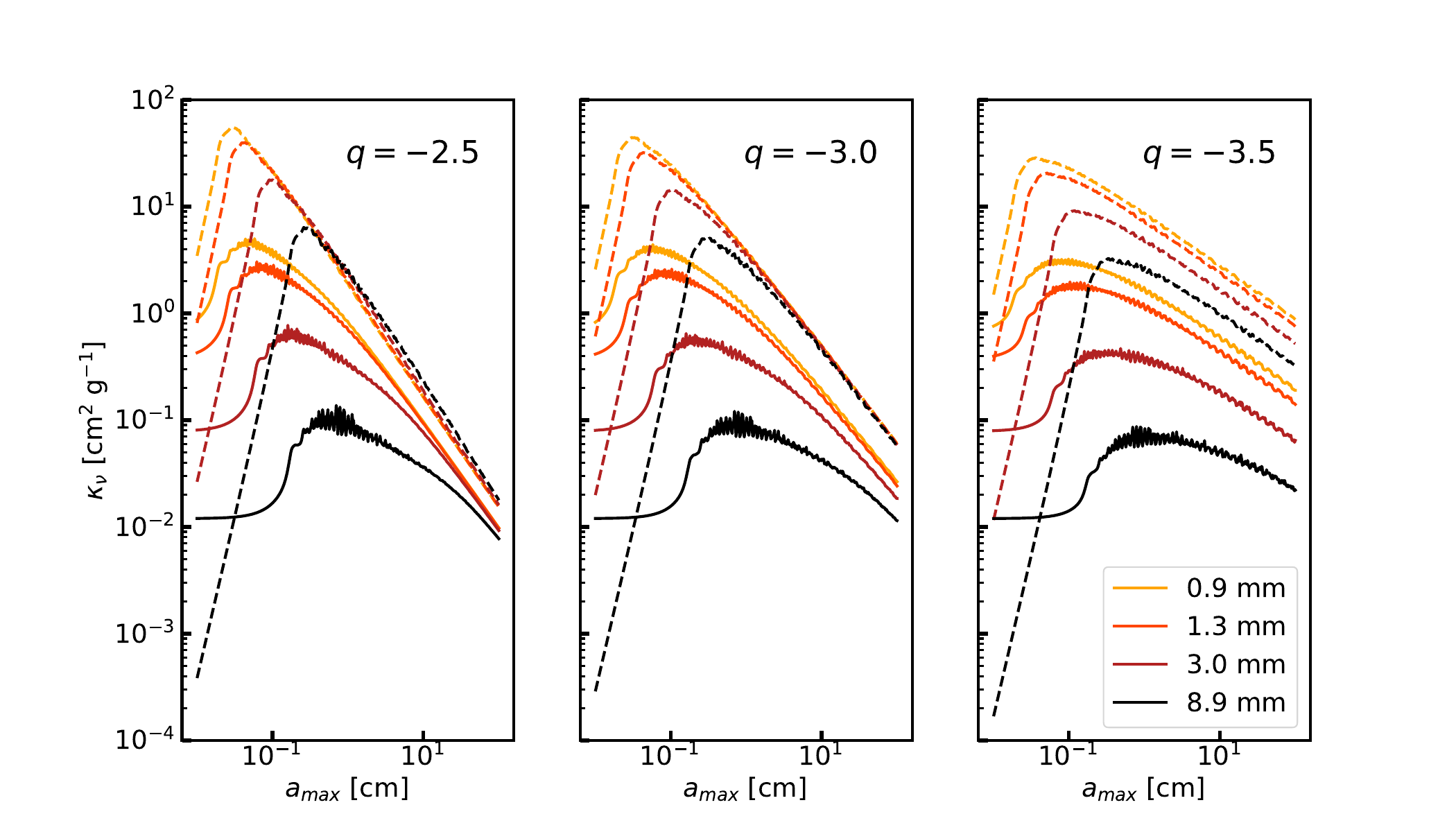}
    \captionof{figure}{Opacities used in this work at each wavelength as a function of the maximum grain size. The solid lines indicate the absorption opacities and the dashed lines the scattering opacities, not corrected from the geometrical factor. The left panel displays the opacities for an exponent of the dust grain size distribution $q=-2.5$, the middle panel for $q=-3.0$, and the right panel for $q=-3.5$.
    The opacities are computed using the DSHARP composition and using the {\sc dsharp\_opac} python package \citep{Birnstiel+2018}. 
}
    \label{fig:opacities}
\end{minipage}
\end{figure*}

Opacities are computed with the {\sc dsharp\_opac} python package \citep{Birnstiel+2018} for a mixture composed by $20\%$ water ice \citep{WarrenBrandt2008}, $33\%$ silicates \citep{Draine2003}, $7\%$ troilite, and $40\%$ of organics \citep{HenningStognienko1996}. The opacities are averaged over the dust size distribution. Figure \ref{fig:opacities} shows the absorption and scattering opacity curves.

\section{Central emission}
\label{app : central_emission}

The intensity maps of V892~Tau on Figure \ref{fig:obs_gallery} show clear emission inside the circumbinary ring. Unresolved emission in the vicinity of the stars could arise from free-free processes (see Section \ref{subsec:sp_ind}) or from a narrow circumstellar disc \citep{Vides+2023}. Our {\sc MCFOST} models did not include a central emission source.
This allows to estimate the signal of the circumbinary dust ring's that was spread to the inner parts due to convolution with the beam. Contamination of the central emission by the circumbinary ring could make the central emission appear brighter and thus bias estimations of the spectral index. We note $F_{in,\text{CBD}}$ the flux of the inner regions coming from the convolution of the ring.
Since the observations were smoothed to a resolution of $0.12\arcsec \times 0.12\arcsec$ that is coarser than their original resolution, our estimates need to be considered as upper limits.
We estimated the contribution of the central emission by the circumbinary ring's signal by integrating the intensity of our smoothed best-fit model (shown in Figure \ref{fig:I_bestfit}, where there is no central emission source) within one beam at the centre of the image. The error on this measurement was calculated by integrating the error on the best-fit radial profile, given by Equation \ref{eq:rad_error}. Table \ref{table:contribution_CBD} summarizes our measurements in mJy and as a fraction of the observed central flux $F_{in}$.
At $8.9$ mm, the convolution with the beam does not spread much signal from the dust ring on the central regions, up to $1\%$ of the observed flux. At $1.3$ mm and $3.0$ mm, the models indicate that up to $67\%$ and $38\%$ of the signal observed in the central regions could actually come from the ring. At $0.9$ mm, the model intensity is overestimated in the inner regions which results in a contribution of the circumbinary ring greater than the observed central flux.
These results indicate that the $F_{in}$ values may be overestimated compared to the actual flux coming from the vicinity of the stars. To measure the impact of this effect on the spectral index, we first defined effective flux values of the inner and outer regions at each wavelength, respectively noted $F'_{in}$ and $F'_{out}$ as:
\begin{eqnarray}
F'_{in} &=& F_{in} - F_{in,\text{CBD}}\,,\\
F'_{out} &=& F_{tot} - F'_{in}\,.
\end{eqnarray}
Using these effective flux values, we found effective spectral indices of $\alpha'_{\text{in}} = 1.62 \pm 0.92$, $\alpha'_{\text{out}} = 2.10 \pm 0.01$ between $0.9$ mm and $3.0$ mm, and of $\alpha'_{\text{in}} = -0.74 \pm 0.19$, $\alpha'_{\text{out}} = 2.77 \pm 0.08$ between $3.0$ mm and $9.8$ mm. Given the close match between these values and the spectral indices derived earlier, our conclusions drawn in Section \ref{subsec:sp_ind} are not affected by the signal spread from the dust ring to the innermost regions by the beam convolution.

\begin{table}
    \centering
    \caption{Contribution of the circumbinary disc to the central emission.}
 \begin{tabular}{c c c} 
Wavelength & \multicolumn{2}{c}{$F_{in,\text{CBD}}$} \\ [0.5ex] 
(mm) & (mJy) & ($F_{in}$)\\
 \hline \hline
$8.9$ & $0.01\pm0.07$ & $0.01\pm0.06$ \\
$3.0$ & $0.3\pm0.2$ & $0.4\pm0.2$  \\
$1.3$ & $2.0\pm0.4$ & $0.7\pm0.1$ \\
$0.9$\tablefootmark{1} & $21\pm3$ & $4.0\pm0.5$ 
\end{tabular}
\tablefoot{$F_{in,\text{CBD}}$ corresponds to the signal of the outer ring spread on the inner regions ($1$ beam area at the center of the image) by the convolution with the beam. These numbers should be taken as upper limits for the datasets used in this work. \\
\tablefootmark{1}{At $0.9$ mm the model intensity is greater than the data intensity.}
}
\label{table:contribution_CBD}
\end{table}


%
\section{Comparison to 1D models}
\label{app : dust_slab}

In the literature, 1D models of dust slab are commonly used to infer the dust properties as a function of radius \citep[e.g.][]{Carrasco-Gonzalez+2019,Sierra+2019}. In this Appendix, we compare our results obtained by a 3D full radiative transfer approach to the results produced by dust slab models. 

Assuming an axisymmetric, vertically thin, and vertically isothermal disc,
the intensity in each radial bin following the model of a 1D vertically isothermal dust slab is \citep{Sierra+2019,Carrasco-Gonzalez+2019} :

\begin{equation}
I_{\nu} = B_{\nu}(T_d) \left[ (1 - \exp(-\tau_{\nu} / \mu)) + \omega_{\nu} F(\tau_{\nu}, \omega_{\nu}) \right],
\end{equation}
with
\begin{equation}
\begin{split}
F(\tau_{\nu}, \omega_{\nu}) = &\frac{1}{\exp(-\sqrt{3} \epsilon_{\nu} \tau_{\nu})(\epsilon_{\nu} - 1) - (\epsilon_{\nu} + 1)}\\
&\times \left[ \frac{1 - \exp\left( -(\sqrt{3} \epsilon_{\nu} + 1 / \mu) \tau_{\nu} \right)}{\sqrt{3} \epsilon_{\nu} \mu + 1}\right.\\
&+ \left.\frac{\exp(-\tau_{\nu} / \mu) - \exp(-\sqrt{3} \epsilon_{\nu} \tau_{\nu})}{\sqrt{3} \epsilon_{\nu} \mu - 1} \right],
\end{split}
\end{equation}
with the individual terms computed as 
\begin{eqnarray}
\tau_{\nu} &=& \Sigma_d (\kappa_{sca,\text{eff}}+\kappa_{abs}) \, ,\\
\mu &=& \cos(i),\\
\epsilon_{\nu} &=& \sqrt{1-\omega_{\nu}} \, ,\\
\omega_{\nu} &=& \kappa_{sca,\text{eff}}/(\kappa_{sca,\text{eff}}+\kappa_{abs}) \, ,\\
\kappa_{sca,\text{eff}} &=& (1-g_{\nu})\kappa_{sca} \, ,
\end{eqnarray}

with $I_{\nu}$ being the intensity at the frequency $\nu$, $B_{\nu}$ the blackbody intensity, $\tau_{\nu}$ the optical depth, $\mu$ the inclination parameter, $\omega_{\nu}$ the albedo of the dust, $\kappa_{sca}$ and $\kappa_{abs}$ the dust opacity coefficients respectively of the absorption and scattering. We followed the approximation of \cite{Carrasco-Gonzalez+2019} for anisotropic scattering by considering an anisotropic scattering coefficient $\kappa_{sca,\text{eff}}$ depending of the forward scattering parameter $g_{\nu}$. The opacity curves were the same as described in Section \ref{subsec:dust_modelling} and shown in Figure \ref{fig:opacities}. The parameter space was defined by values of surface density between $10^{-2}$ and $10^2$~g~cm$^{-2}$, values of temperature between $1$ and $350$~K, and values of the maximum grain size between $10^{-2}$ and $10^2$~cm. We fixed $q=-3.5$. Each parameter was sampled over a grid of $400$ bins, which were logarithmically spaced in the case of the surface density and maximum grain size. We explored the parameter space using an MCMC sampling implemented in {\sc emcee} \citep{emcee} and a Gaussian likelihood $\mathcal{L}\propto\exp(-\chi^2_{\text{dust,1D}}/2)$ with $\chi^2_{\text{dust,1D}}$ defined as in Equation \ref{eq:chi2_dust}. We initialized $24$ walkers each going for $20000$ steps after a burn-in phase of $1000$ steps. Uniform priors were chosen to follow the input parameter space. Following the approach of \cite{Macias+2021}, the temperature was initialized with a Gaussian prior with mean value the radial temperature of a passively irradiated flared disk in radiative equilibrium \citep{ChiangGoldreich1997,Dullemond+2001} : 

\begin{equation}
\label{eq:prior_T}
T(r) = \left( \frac{\varphi L_*}{8\pi r^2 \sigma_{\text{sb}}} \right)^{0.25},
\end{equation}

with $\varphi$ being the flaring angle of the disc, $L_*$ the stellar luminosity, and $\sigma_{\text{sb}}$ the Stefan-Boltzmann constant. For the flaring angle, we used a classical value of $\varphi = 0.35 \pm 0.25$. For the stellar luminosity, we assumed $L_* = 144$ L$_{\odot}$ based on the stellar models of \cite{Siess+2000} for two $3$ M$_{\odot}$ at $3$ Myr consistently with the age of the system \citep{KucukAkkaya2010}. We arbitrarily assumed an error of $10\%$ for $L_*$. 
We performed a first run in these conditions (which we call the general slab model), and a second run where we imposed a prior on the optical depth at $8.9$~mm such that $\tau_{\text{8.9mm}}<1$. The dust slab models are not fitted in the regions where the binary has been subtracted in the VLA data, and in the outer regions if the intensity is not monotonously decreasing with wavelength.
To establish a proper comparison with the {\sc MCFOST} models, we performed the same analysis than in Section \ref{subsec:dust_modelling} but fixing $q=-3.5$ and including the $0.9$~mm data in the fitting process. The observations and {\sc MCFOST} models were smoothed to the resolution of that dataset, namely $0.19\arcsec\times0.13\arcsec$.

\begin{figure*}
\centering
\begin{center}
    \includegraphics[width=\textwidth, trim={0cm 0cm 0cm 0cm},clip]{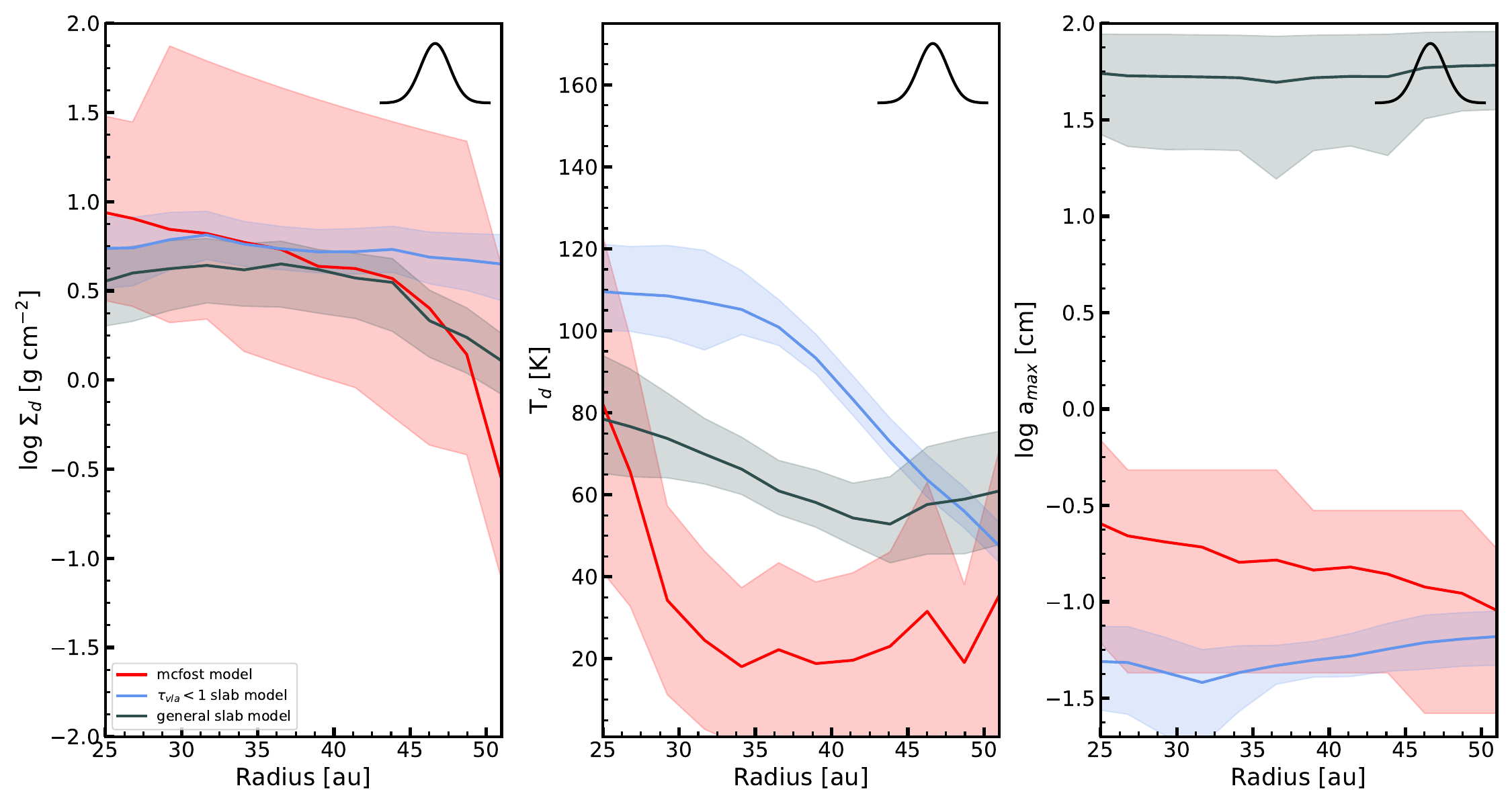}  
    \caption{Marginalized probability distribution resulting from the fitting of radiative transfer models to the data radial profiles. The red curve corresponds to the expectation of the {\sc MCFOST} computed with Equation \ref{eq:expectation}, the blue curve to the one of the dust slab model including an optically thin prior at $8.9$~mm, and the grey curve to the one dust slab model without the optical depth prior. The shaded area corresponds to the errors, taken as $1\sigma$ confidence interval for the {\sc MCFOST} models, and delimited by the $16$th and $84$th percentiles for the slab models. 
    The inset at the top right of each panel represents the synthesized beam of the observations. 
}
    \label{fig:slab_post}
\end{center}
\end{figure*}

For the case with the prior on optical depth, the walkers achieved convergence with auto-correlation lengths of $57$, $61$, and $63$ for $\Sigma_d$, $T_d$, and $a_{max}$, respectively. For the case without the optical depth prior, these same auto-correlation lengths are $116$, $87$, and $100$. Ignoring the first $5000$ steps, the marginalized posterior distributions resulting of the MCMC sampling can be found in Figure \ref{fig:slab_post}. We also show on that Figure the results obtained with {\sc MCFOST}. The plain curves correspond to the expectation computed with Equation \ref{eq:expectation}. The errors are calculated using the $1\sigma$ confidence intervals for {\sc MCFOST}, and the $16$th and $84$th percentiles of the marginalized posterior distributions for the slab models. 
The results for the dust surface density are consistent with each other, but discrepancies are observed for the dust temperature and maximum grain size. 
{\sc MCFOST} temperatures displayed here are mass-averaged, which means that it traces dust temperature close to the dense midplane. Since optically thick observations are included in the modelling, and that the slab models fits the temperature as a free parameter, the temperature profile found by slab models by construction traces the temperature of the $\tau=1$ layer of the optically thick observations. That layer is possibly elevated compared to the protected dense midplane, and so hotter.
The maximum grain size is not constrained in the case of the general slab model, with the walkers stacking at the higher bound of the parameter space. The slab model forced optically thin at $8.9$~mm finds maximum grain sizes of approximately $0.05$~cm, which is compatible with what we find with {\sc MCFOST}.

\begin{figure*}
\centering
\begin{center}
    \includegraphics[width=\textwidth, trim={0cm 0cm 0cm 0cm},clip]{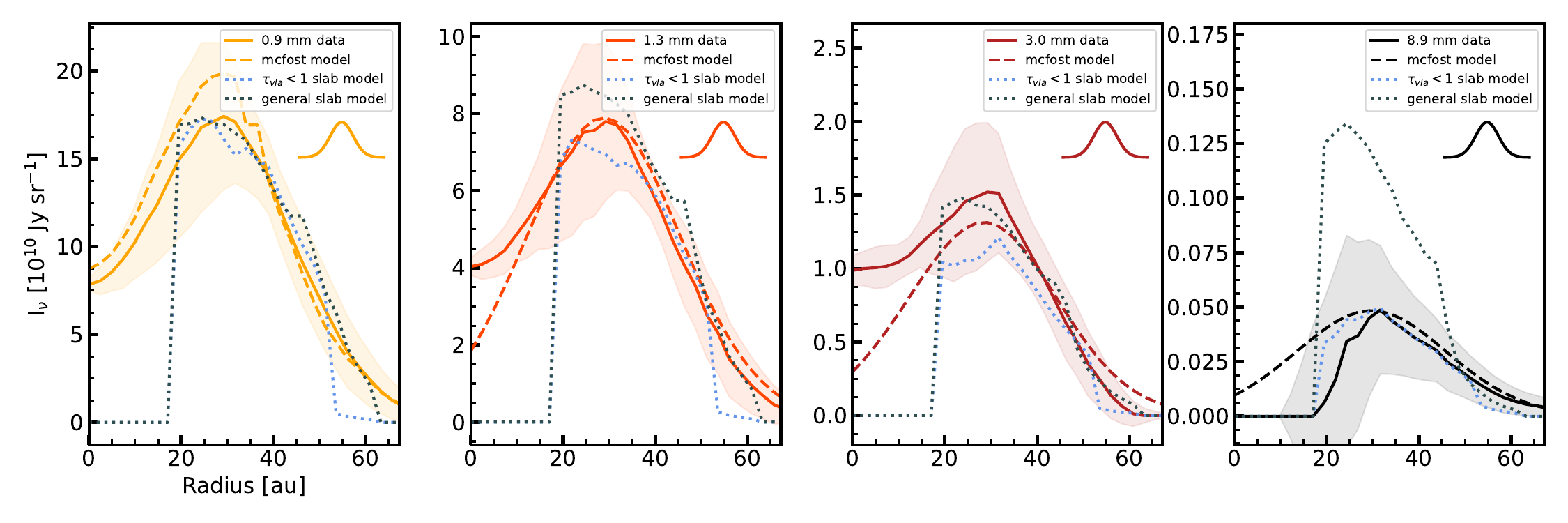}  
    \caption{Same as Figure \ref{fig:I_bestfit} but the best-fit intensity radial profiles of the 1D models are indicated by the dotted lines at each wavelength. The light blue and dark blue lines corresponds to the models with the $\tau<1$ at $8.9$~mm and without, respectively. The data and best-fit {\sc MCFOST} models at $1.3$~mm and longer were convolved to the resolution of the $0.9$~mm image to match the resolution of the slab models.   
}
    \label{fig:slab_bestI}
\end{center}
\end{figure*}

Figure \ref{fig:slab_bestI} shows the best-fit intensity profiles of the 1D and 3D models alongside the data radial profiles at each wavelength. The radial profiles of the 1D model including an optically thin prior fall within the $1\sigma$ error at all the wavelength and between $25$ and $51$~au, which is the extent of the 3D model. The same can be said for the slab model without the prior up to $3.0$~mm, beyond which the intensity is largely overestimated. 

The mismatch between the general slab model and the data at $8.9$~mm likely comes from difficulties encountered by the fit to constrain the optical depth. This is illustrated by the walkers being systematically limited by the upper limit of the parameter space in maximum grain size (or in surface density in some other runs). Poorly constrained optical depths could be a consequence of the lack of fully optically thin data in the set of observations. It also suggests that the DHSARP opacities are not the most appropriate ones to reproduce the SED at each radius. Repeating the methods of this Section with opacities of different dust composition would likely provide a better fit to the data and give insights on the dust composition in V892~Tau. We let that study to be explored in future works.


\section{Correlations between radial bins}
\label{app : unique_mcfost_model}

\begin{figure*}
\centering
\begin{center}
    \includegraphics[width=\textwidth, trim={0cm 0cm 0cm 0cm},clip]{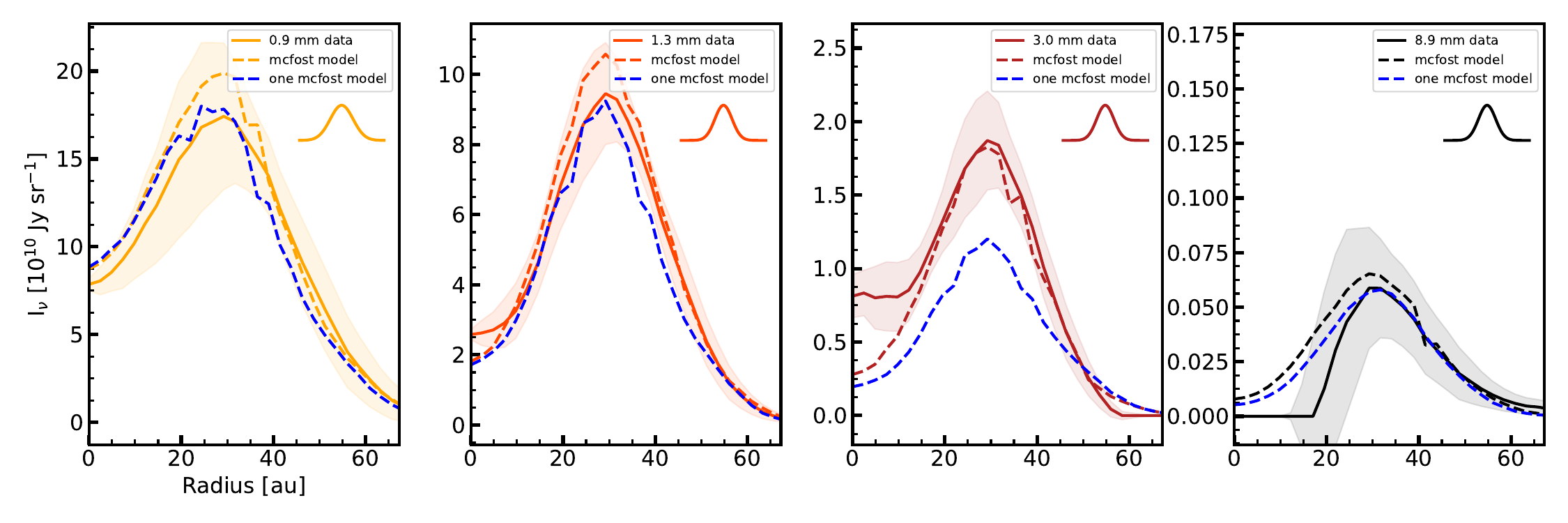}  
    \caption{Same as Figure \ref{fig:I_bestfit} but the intensity radial profiles of the unique {\sc MCFOST} model best matching the data are indicated by the dotted blue line at each wavelength.    
}
    \label{fig:I_profile_correlations}
\end{center}
\end{figure*}

Our methods in modelling the dust properties in V892~Tau involved convolving images of radiative transfer models with the observed beam and then perform a fitting procedure independently at each radius (see Section \ref{subsec:dust_modelling}). For a given radial bin, the convolution with the beam introduces a correlation between the intensity in the neighbouring radial bins. These correlations are not taken into account by our fitting method. To assess the impact of this effect on our results, we constructed a single radiative transfer model using our best-fit parameters. This unique {\sc MCFOST} model is defined by the surface density distribution, the maximum grain size, and the power-law exponent of the grain size distribution that together provide the best match to the observed intensity data (see Figure \ref{fig:I_bestfit}). The other input parameters remained defined as described in Section \ref{subsec:dust_modelling}. The output images were convolved the images of the model at $1.3$~mm, $3.0$~mm, and $8.9$~mm with a $0.12\arcsec$ circular beam. The image at $0.9$~mm was convolved with the same beam as the observation at the corresponding wavelength. Figure \ref{fig:I_profile_correlations} shows the radial profiles of the unique {\sc MCFOST} in comparison to the best fit profiles found by our exploration of the parameter space.

Except at $3.0$~mm, the radial profiles of the single {\sc MCFOST} model reproduce well the data and show a comparable agreement with the radial profiles derived in Section \ref{subsec:dust_modelling}. The discrepancy at $3.0$~mm indicate that, at that wavelength, radial correlations play an important role in the resulting intensity radial profiles. To match the intensity at $3.0$~mm, greater $\Sigma_d$ and $T_d$ would have been needed. However given the close match single {\sc MCFOST} model for the other datasets, we argue that the input parameters are still well constrained.
This stresses that the correlation between the radial bins does not play a significant role in the spatial distribution of the dust properties, as modelled in our work.

\end{appendix}

\end{document}